\documentclass[twocolumn,openany,openright,amsmath,amssymb,superscriptaddress,prb]{revtex4-2}

\usepackage{float}
\usepackage{latexsym} 
\usepackage{amsmath,amsthm}
\usepackage{ifpdf}
\usepackage{epstopdf}
\usepackage{soul}
\usepackage{dcolumn}
\usepackage{bm}
\usepackage{braket}
\usepackage{wrapfig}
\usepackage[abs]{overpic}
\usepackage{graphicx,graphics}
\usepackage{dcolumn}
\usepackage{wasysym}
\usepackage{latexsym,verbatim}
\usepackage{color}
\usepackage[framemethod=default]{mdframed}
\usepackage[breaklinks=true,colorlinks,citecolor=blue,linkcolor=blue,urlcolor=blue]{hyperref}
\usepackage[makeroom]{cancel}
\usepackage{fancybox}

\newcommand{\upp}{\uppercase}

\newcommand {\Pdert}{\frac {\partial}{\partial t}}

\newcommand{\sx}{\sigma_{\rm x}}
\newcommand{\sy}{\sigma_{\rm y}}
\newcommand{\sz}{\sigma_{\rm z}}

\newcommand{\kk}{\mathbf{k}}

\newcommand{\pp}{\mathbf{p}}
\newcommand{\qq}{\mathbf{q}}

\newcommand{\rr}{\mathbf{r}}

\newcommand{\AAA}{\mathbf{A}}

\begin{document}
\def \k{\bold k}
\def \l{\bold l}
\def \i{\bold i}
\def \p{\bold p}
\def \r{\bold r}
\def \qq{\bold q}
\def \A{\bold A}
\def \beq{\begin{equation}}
\def \eeq{\end{equation}}
\def \beal{\begin{aligned}}
\def \eal{\end{aligned}}
\def \bes{\begin{split}}
\def \ees{\end{split}}
\def \besu{\begin{subequations}}
\def \esu{\end{subequations}}
\def \g{\gamma}
\def \G{\Gamma}
\def \ac{\alpha_c}
\def \barr{\begin{eqnarray}}
\def \earr{\end{eqnarray}}

\title{Optotwistronic of bilayer graphene}
\author{Leone Di Mauro Villari}
\email{leone.dimaurovillari@manchester.ac.uk}
\author{Alessandro Principi}
\affiliation{Department of Physics and Astronomy, University of Manchester, Manchester M13 9PL, UK}
\begin{abstract}
We present a study of the nonlinear optical response of twisted bilayer graphene. We discuss the contribution of the Berry phase to the non-linearity when  inversion symmetry is broken, thus underlining the interplay between band and real space geometry, and nonlinear response. We also highlight an effect which is characteristic of extreme nonlinear optics: the generation of harmonics in disguise. This effect emerges in twisted bilayer graphene at relatively moderate fields strengths because of the much reduced band width. Our findings contribute to the understanding of the link between geometry and optical properties, as well as of the extreme nonlinear optical regime in twisted bilayer graphene. 
\end{abstract}
 
\maketitle
\section{Introduction}
 The  discovery  of  correlated  phases  for Dirac-like  electrons in twisted  bilayer  graphene (TBG) has  paved  the  way  for  a  large  amount of  research  on the  relation  between  the  geometry  of  a  lattice  and its electronic  properties \cite{nim}. Of particular interest is the emergence of flat bands at specific twisting angles (magic angles). In this case, TBG becomes superconductive and exhibits correlated-insulating phases at integer filling fractions \cite{cao,xie}. Hitherto, a number of microscopic theories have been developed to understand such new phenomena, concerning not only unconventional superconductivity but also correlated insulation 
 \cite{BiMac,ssk,pm,pla,kmn} (see also~\cite{He2021,Andrei2021} and references therein). 
On the contrary, the nonlinear optical response of TBG has started to draw some attention only recently. Floquet band theory has been proposed as a method to tune magic angles and in general to get control over the 
twisted-graphene physics by modifying intra- and inter-layer hopping amplitudes with a driving field \cite{tj,vm,vm1}. The photogalvanic effect has also been investigated by means of perturbative methods and the Boltzmann equation \cite{oh,gy}. On the contrary, only a handful of studies focuses on a proper theoretical description of the harmonic generation process \cite{ike,mcz,zz,sn}. 

In this paper we use a non-perturbative approach based on the formalism of the Dirac Bloch equations (DBEs) \cite{Ishikawa_2010,cb,cm,vg} to study the high-order response of TBG and how such response varies with the twisting angle. We elucidate the contribution to the nonlinear current of intraband and interband transitions. Furthermore, we study the variation of the current spectra due to the introduction of an energy gap. In particular we show that a complex interplay between lattice geometry (twisting) in real space, eigenstate geometry (Berry phases) in momentum space and optical response emerges naturally from the dynamical equation. This topic has attracted significant attention recently and has been mostly dealt within perturbation theory \cite{jm,mh}. We also highlight the phenomenon of even harmonics in disguise which is peculiar of extreme nonlinear optics. 

The DBEs are based on the formalism of instantaneous eigenstates. These equations parallel the well known semiconductor Bloch equations \cite{lk}, but they are non-perturbative and encapsulate both the intraband and interband dynamics. They have been introduced for the first time by Ishikawa in 2010 \cite{Ishikawa_2010} to study the nonlinear response of graphene and later extended to include doping effects \cite{cb}, gap opening \cite{cm} and Coulomb interactions \cite{vg}. More recently they have also been applied to materials presenting Type-II (tilted) Weyl low energy dispersion \cite{tvo1}. This paper is organized as follows. In the next section we review the formalism of instantaneous eigenstates, introduced in \cite{Ishikawa_2010}, for the simple case of a two bands model. In section III we briefly review the continuum limit of TBG, we introduce the electromagnetic interaction and derive the DBEs. In section IV we study the nonlinear response to a short electromagnetic laser pulse in two different configurations: gapless flat bands at the magic angle, and gapped bands both away and at the magic angle. In this last configuration we study the generation of odd harmonics \emph{in disguise} of even harmonics.

\section{Instantaneous eigenstates formalism}
\label{sect:IEF}
In this section we review the application of the instantaneous eigenstates formalism to the case of a two-band model. Traditionally this approach is used to describe time-dependent Hamiltonian quantum systems in the adiabatic limit, i.e. under the assumption that the system does not do transitions from an instantaneous eigenstate to another during a long time interval $t$ ($t \to \infty$ in the adiabatic limit) \cite{sakurai_napolitano_2020}. We will see that, in our case, such assumption is not necessary \cite{Ishikawa_2013}. 

We start from the following time dependent Schr\"odinger equation in momentum space
\beq \label{dk}
i \Pdert \psi_{\kk}(t) =  H_\kk(t) \psi_\kk(t).
\eeq 
 with $H_\kk(t) = H_{\kk + \AAA(t)}$, where $\AAA(t)$ is a homogeneous external vector potential. We make the following ansatz for the solution
 \beq \label{ista}
  \psi_{\kk}(t) = \sum_{\lambda} c^\lambda_{\kk}(t) \varphi^\lambda_{\kk}(t) e^{- i \mathcal E^\lambda_{\kk}(t)}
 \eeq
 where $c^\lambda_{\kk}(t)$ are expansion coefficients, $\lambda=\pm1$ is the band index and $\mathcal {E}^\lambda_{\kk}(t)$ is a time dependent phase to be determined. The states $\varphi^\lambda_{\kk}(t)$  are the so called instantaneous eigenstates which are an exact solution of the instantaneous eigenvalue problem 
 \beq \label{ista2}
 H_\kk(t) \varphi^\lambda_{\kk}(t)  = \epsilon^\lambda_{\kk}(t) \varphi^\lambda_{\kk}(t).
 \eeq
 Here, $\epsilon_{\kk}^\lambda(t)$ is the instantaneous eigenvalue. Substituting equation (\ref{ista}) into (\ref{dk}) we rewrite its left-hand side as (from now on, we drop the explicit time dependence in longer expressions)
 \beq \label{dk1}
 i \Pdert  \psi_{\kk} = i \sum_{\lambda} ( \dot c^\lambda_{\kk}  \varphi^\lambda_{\kk}  +  c^\lambda_{\kk}   {\dot \varphi}^\lambda_{\kk} -i c^\lambda_{\kk} \dot{ \mathcal E}^\lambda_{\kk}  \varphi^\lambda_{\kk}  ) e^{-i \mathcal E^\lambda_{\kk}}. 
\eeq
As for the right-hand side of equation (\ref{dk}), using equation (\ref{ista2}) we obtain 
 \beq \label{dk2}
 H_{\kk} \psi_{\kk} =   H_{\kk} \sum_{\lambda} c^\lambda_{\kk} \varphi^\lambda_{\kk}e^{- i \mathcal E^\lambda_{\kk}} = \sum_{\lambda} \epsilon^\lambda_{\kk} \,c^\lambda_{\kk} \varphi^\lambda_{\kk}e^{-i \mathcal E^\lambda_{\kk}},
 \eeq
the compatibility between~(\ref{dk1}) and~(\ref{dk2}) can be realized by defining the \emph{dynamical phase} 
 \beq
 \mathcal E^\lambda_{\kk}(t) = \int_{-\infty}^t \epsilon^\lambda_{\kk}(t') \, dt',
 \eeq
and simultaneously eliminating the first two terms on the right hand-side of equation (\ref{dk1}). 
We consider the equation
\beq
 [\dot c^\lambda_{\kk}(t)  \varphi^\lambda_{\kk}(t)  +  c^\lambda_{\kk}(t)   {\dot \varphi}^\lambda_{\kk}(t)]e^{i \mathcal E^\lambda_\kk(t)} = 0
\eeq
and we multiply it by $ \varphi^{\bar \lambda,*}_{\kk}(t)$. Then summing over $\bar \lambda=\pm \lambda$ and using the state orthonormality ($\varphi^{\bar \lambda,*}_{\kk}(t) \cdot \varphi^\lambda_{\kk}(t) =\delta_{\lambda,{\bar \lambda}}$), we get  
\beq \label{evo}
\dot c^\lambda_{\kk} = i\Omega_{\kk} \,c^{-\lambda}_{\kk} e^{i \mathcal E^\lambda_\kk - i \mathcal E^{-\lambda}_\kk} + i \dot \gamma_\kk \, c^\lambda_{\kk},
\eeq
where we have defined the two quantities $ \dot \gamma_\kk(t) = i  \varphi^{\lambda,*}_{\kk}(t) \cdot {\dot \varphi}^\lambda_{\kk}(t)$ and $\Omega_\kk(t)=-i\varphi^{\bar \lambda,*}_{\kk}(t) \cdot  {\dot \varphi}^\lambda_{\kk}(t)$ (the Rabi frequency -- see below). We observe that the second term in equation (\ref{evo}) can be removed by a local gauge transformation of the wavefunction as 
\beq
 \psi_{\kk}(t) \to  \psi_{\kk}(t) e^{i \gamma_\kk(t)}
\eeq
where $\gamma_\kk(t) = \int_{-\infty}^t  \dot \gamma_\kk(t') dt'$ is a time-dependent Berry phase. The time variation equation (\ref{evo}) thus reduces to
\beq \label{evo1}
\dot c^\lambda_{\kk}(t) = i\Omega_{\kk}(t) \,c^{-\lambda}_{\kk}(t) e^{i \mathcal E^\lambda_\kk(t) - i \mathcal E^{-\lambda}_\kk(t)}.
\eeq
As we show in equation (\ref{var}) below, equation (\ref{evo1}) is used to derive the DBEs.

equation (\ref{evo1}) plays an important role in the theory of adiabatic evolution of quantum systems.
In the proof of the adiabatic theorem, this equation corresponds to requiring adiabaticity. In fact, as shown by Ishikawa in Ref. \onlinecite{Ishikawa_2013}, equation (\ref{evo1}) admits solutions in both the adiabatic and the diabatic limit.  They considered the case of graphene (massless Dirac fermions)  when the electron momentum varies along a circular path around the Dirac point. This situation can be realised under normal incidence of a circularly polarized pulse in the linear regime. In this case equation (\ref{evo}) is analytically solvable and it describes two different dynamics in the adiabatic and diabatic limit. In the first case the electron remains in the state fixed by the initial condition. If for example $c_{\kk,1}=1$ and $c_{\kk,-1}=0$, then it will remain in the upper band. At the same time the instantaneous wave-function acquires a constant Berry phase $\pi$ when the electron completes a cycle. In the diabatic limit instead the electron population is completely transferred to the lower band at half a cycle and it is transferred back to the upper one after a cycle. In contrast to the adiabatic limit the Berry phase is cancelled by a phase acquired through the interband dynamics. These considerations can also be applied to a gapped material (massive Dirac fermions) in the linear optical regime. Interestingly, it has been shown that for massive Dirac fermions in the nonlinear regime, the impact of the Berry phase on the low momentum state dynamics is not negligible even for short time intervals ({\it i.e.} in the diabatic limit) \cite{cm}.

\section{The model}
\begin{figure}[htbp]
         \includegraphics[width=\linewidth]{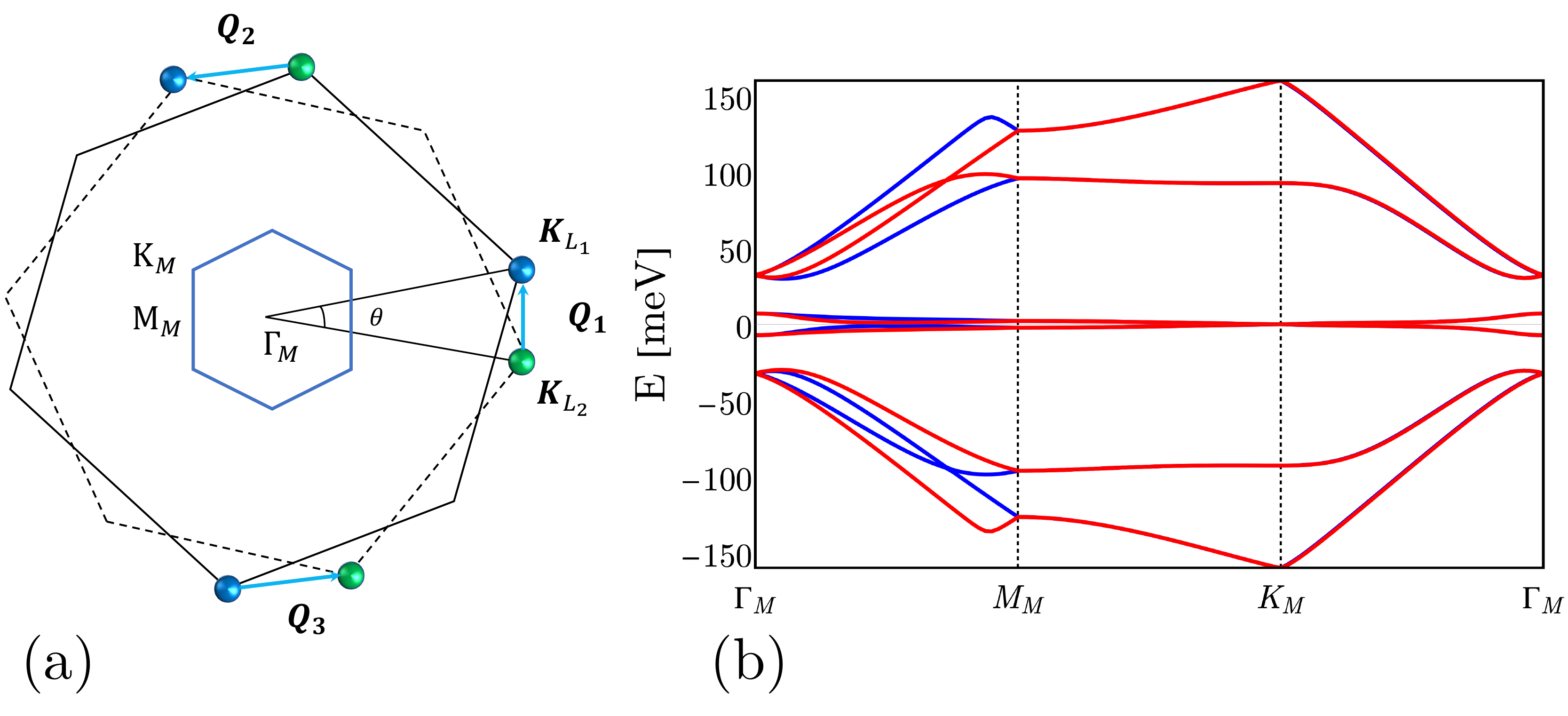}

    \caption{(a) Momentum space geometry of TBG, the small central hexagon is the BZ of the moir\`e superlatice, $\Gamma_M$, $M_M$ and $K_M$ are the high symmetry points. The larger hexagons represent the BZs for the upper and lower graphene layers. (b) Band structure for the six lowest energy bands in the $K$ (red) and  $K'$ (blue) valleys.}
    \label{fig1}
\end{figure}

We begin by introducing the lattice structure and the model Hamiltonian we use as a starting point of this work. We consider two layers of graphene with a modulated mismatch in the relative position of the two lattices of a bilayer, obtained by twisting the upper (lower) layer by an angle $\theta/2$ $(-\theta / 2)$. The resulting mismatch produces a characteristic moir\`e pattern. In the low energy limit $\epsilon \leqslant 1$ eV this system can be described by the following Hamiltonian \cite{BiMac,gg}
\beq \label{hTBG}
 H(\kk) = \left(\begin{array}{cc} H^+_D(\kk) & \hat T^\dagger(\rr) \\ \ \hat T(\rr) &H^-_D(\kk) \end{array}\right).
\eeq
$H^{\pm}(\kk)= v_F \boldsymbol{\sigma}_{||}\cdot(\kk + \pm \Delta\mathbf{K}/2) + \sz \Delta_M/2$ are the single layer graphene  Hamiltonians, here $\Delta_M$ is an energy gap at the Dirac point of the graphene monolayers, due to broken inversion symmetry, $\Delta \mathbf K$ is the shift in the relative position of the Dirac points in the two layers and $\boldsymbol{ \sigma}_{||}=(\sx,\sy)$. The hopping matrix $\hat T(\rr)$ represents the interlayer hopping amplitude, which reflects the spatial alternation of the stacking configuration, ($AA'$, $AB'$ and $BA'$) due to the moir\`e pattern. Here $A$ ($A'$) and $B$ ($B'$) correspond to the two sublattices of the lower (upper) layer, respectively. As usual \cite{BiMac,gg}, we assume that interlayer hopping is dominated by processes with momentum transfer $\mathbf{Q}_0=0$ and $\mathbf{Q}_{12}=(\pm 2\pi/\sqrt 3,2\pi)$ (figure \ref{fig1}a) so that we can write the hopping matrix  elements as $\hat T_{lm}=\sum_{j} u_{_{lm}}\, e^{i\mathbf Q_j \rr}$, $(l,m)$ being layer-sublattice indices. 
We expressed the Hamiltonian (\ref{hTBG}) as a $4N \times 4N$ matrix in $k$-space by using a plane wave expansion, with $N=60$ being the number of plain waves, and we diagonalised it numerically (figure \ref{fig1}b). The eigenvalues and eigenstates obtained from the Hamiltonian expanded in plane waves constitute the set up for the study of the system coupled to the electromagnetic radiation. 

We now introduce the coupling with an impinging electromagnetic field. The minimally coupled plane-wave Hamiltonian reads 
\beq \label{pw}
H_{\kk,\qq_i,\qq_j}(t)= [h^{(0)}_{\kk+e\AAA(t),\qq_i,\qq_j}+ V^{tw}_0]\delta_{\qq_i,\qq_j} + V^{tw}_{\qq_i-\qq_j}
\eeq
where $h^{(0)}_\kk$ is the uncoupled double-layer Hamiltonian and $V^{tw}$ is the plane wave expansion of the twisting potential matrix. Hence, we can write the related time dependent \emph{Dirac} equation for the low energy TBG Hamiltonian wavefunction
\beq \label{Dirac}
i\partial_t \psi_{\kk,\qq_i}(t) = \sum_{\qq_j} H_{\kk,\qq_i,\qq_j} \psi_{\kk,\qq_i}(t)
\eeq
the solution of this equation is obviously rather complicated but, in analogy to what shown in Sect.~\ref{sect:IEF}, it can be expressed as a superposition of instantaneous eigenstates which diagonalize the interacting time-dependent Hamiltonian (\ref{pw}). Using steps analogous to those shown in Sect.~\ref{sect:IEF}, we obtain
\beq \label{spinor}
\psi_{\kk,\qq_i}(t) = \sum_{\lambda}c^\lambda_{\kk}(t)\varphi^\lambda_{\kk,\qq_i}(t)e^{-i \gamma^\lambda_{\kk}(t) -i\mathcal E^\lambda_\kk(t)}.
\eeq
Here, $\lambda$ is the band index, $\varphi^\lambda_{\kk,\qq_i}(t)$ and $\epsilon^\lambda_\kk(t)$ are instantaneous band eigenstates and eigenvalues which solve the following eigenvalue problem
\beq
\sum_{\qq_j} H_{\kk,\qq_i,\qq_j}(t)\varphi^\lambda_{\kk,\qq_j}(t)= \epsilon^\lambda_\kk(t) \varphi^\lambda_{\kk,\qq_i}(t).
\eeq
The extra phase term is the Berry phase which, in analogy with Sect.~\ref{sect:IEF}, is defined as 
\beq
\gamma^\lambda_\kk(t) =\sum_{\qq_i} \int_{-\infty}^t \varphi^{\lambda,\dagger}_{\kk,\qq_i}(t)\dot \varphi^{\lambda}_{\kk,\qq_i}(t),
\eeq
where in this equation we have used hermitian conjugation as the instantaneous plane wave eigenstates $\varphi^{\lambda}_{\kk,\qq_i}(t)$ are four-components spinors. By substituting equation (\ref{spinor}) in equation (\ref{Dirac}) we can derive a system of coupled differential equations, the \emph{Dirac-Bloch} equations, for the population inversion and microscopic polarisation 
\beq \label{var}
\begin{aligned}
w^{\lambda,\lambda'}_{\kk}(t)&=|c^\lambda_\kk(t)|^2-|c^{\lambda'}_\kk(t)|^2,\\ p^{\lambda,\lambda'}_\kk(t)&=c^\lambda_\kk(t)c^{*\lambda'}_\kk(t)e^{-i [\mathcal E^\lambda_\kk(t')-\mathcal E^{\lambda'}_{\kk}(t')]}. \end{aligned}
\eeq
The resulting system is numerically quite demanding as it comprises of a set of $4N(4N-1)$ coupled differential equations (see Appendix). In what follows we consider the dynamics of the lowest energy bands only, this allows to have a clear qualitative picture of the nonlinear response without having to solve an excessively large system. For the case of two bands the DBEs read
\beq \label{RDB}
\begin{cases}
\dot p_{\kk} &= -i[\omega_0-\delta\epsilon_{\kk}(t)] p_{\kk} - i \Omega_{\kk}(t)\,e^{-i\delta\gamma_{\kk}(t)+i\omega_0 t} w_\kk, \\
\\
\dot w_{\kk} &= - 4\operatorname{Re}\left\{\left(\Omega_{\kk}(t)\right)^*\,e^{i\delta\gamma_{\kk}(t)+i\omega_0t}p_{\kk}\right\},
\end{cases}
\eeq
where $\omega_0$ is the central frequency of the impinging field and $\delta \epsilon_\kk(t)$ ($\delta \gamma_\kk(t)$) is the energy (Berry phase) difference between the two lowest energy bands. The quantity $\Omega_{\kk}(t)$ is the Rabi frequency of the interacting system and is defined as 
\beq
\Omega_{\kk}(t) = -i\boldsymbol \mu_\kk(t) \cdot \mathbf E(t) = -i\sum_{\qq_i} \varphi^{c,\dagger}_{\kk,\qq_i}(t)\dot \varphi^{v}_{\kk,\qq_i}(t)
\eeq
where $ \boldsymbol \mu_\kk(t)=\boldsymbol \mu_{\kk + e \AAA(t)}$ is the time dependent dipole moment, $\mathbf E(t)=-\dot \AAA(t)$ is the impinging electric field and $c$ ($v$) denotes the conduction (valence) band. To characterize the nonlinear response of the system from the solution of the DBEs we compute the time dependent optical current which is defined as 
\beq
J^{\mu}(t) =-e\sum_{\qq_i,\qq_j,\kk}\, \psi^\dagger_{\kk,\qq_i}(t)  v^{\mu}_{\kk,\qq_i,\qq_j}\, \psi_{\kk,\qq_i}(t),
\eeq
where $v^{\mu}_{\kk,\qq_i,\qq_j}=\partial_{k_\mu}H_{\kk,\qq_i,\qq_j}$ is the velocity operator. Using equation (\ref{spinor}) and the definition of population and inversion variables we can separate the current into intraband and interband contributions as
\beq \label{curr}
\begin{split}
J^{\mu}(t) = &\sum_\kk \Bigl[\Bigl (J_{\kk,intra}^{\mu,c} - J_{\kk,intra}^{\mu,v}\Bigl)\frac{w_\kk + 1}2 \\ + &J_{\kk,inter}^{\mu} \operatorname{Re}\Bigl(p_\kk e^{-i(\Delta \gamma_\kk(t)+\omega_0t)}\Bigl)\Bigl]
\end{split}
\eeq
where $J_{\kk,intra}^{\mu,\lambda=c,v}=-e\sum_{\qq_i,\qq_j} \varphi^{\lambda,\dagger}_{\kk,\qq_i}(t) v^{\mu}_{\kk,\qq_i,\qq_j} \varphi^{\lambda}_{\kk,\qq_i}(t)$ is the intraband contribution to the nonlinear optical current, while $J_{\kk,inter}^{\mu}=-e\sum_{\qq_i,\qq_j} \varphi^{c,\dagger}_{\kk,\qq_i}(t) v^{\mu}_{\kk,\qq_i,\qq_j} \varphi^{v}_{\kk,\qq_i}(t)$  is the interband one. Note that to simplify the notation we have written our equation in one $K$ valley of the original double layer Brlloiun zone. In the numerical simulation both valleys have been considered to avoid introducing a spurious time reversal symmetry breaking.
\section{Nonlinear Optical response}
\subsection{Traditional nonlinear optics }
\begin{figure*}[htbp]
        
        \begin{minipage}[b]{.32 \textwidth}
         \centering
         \begin{overpic}[width=\linewidth]{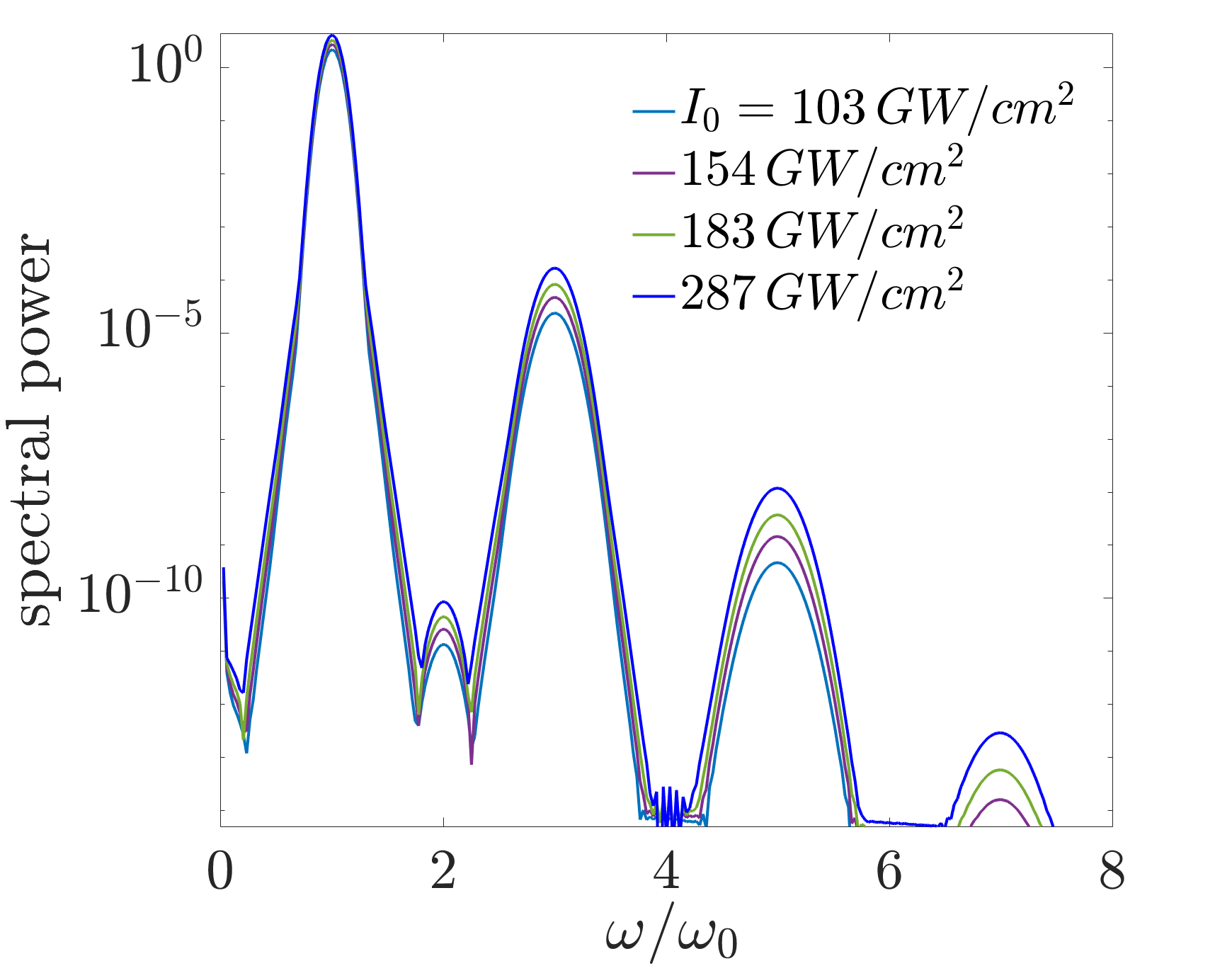}
         \put(0,0){(a)}
    \end{overpic}
     \end{minipage}\hfill
     \begin{minipage}[b]{ .32\textwidth}
         \centering
         \begin{overpic}[width=\linewidth]{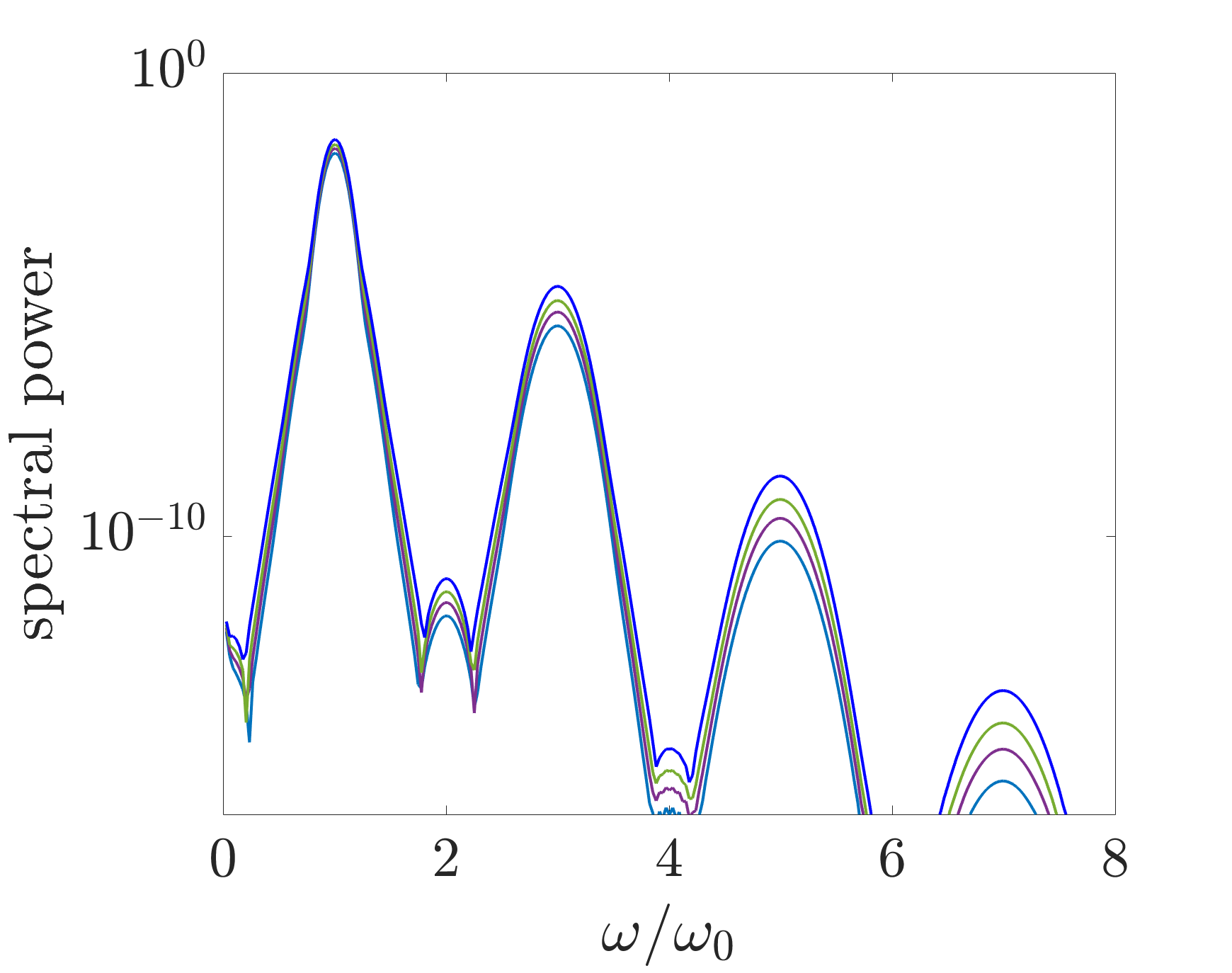}
         \put(0,0){(b)}
    \end{overpic}
     \end{minipage}\hfill
     \begin{minipage}[b]{ .32\textwidth}
         \centering
    \begin{overpic}[width=1.0\linewidth]{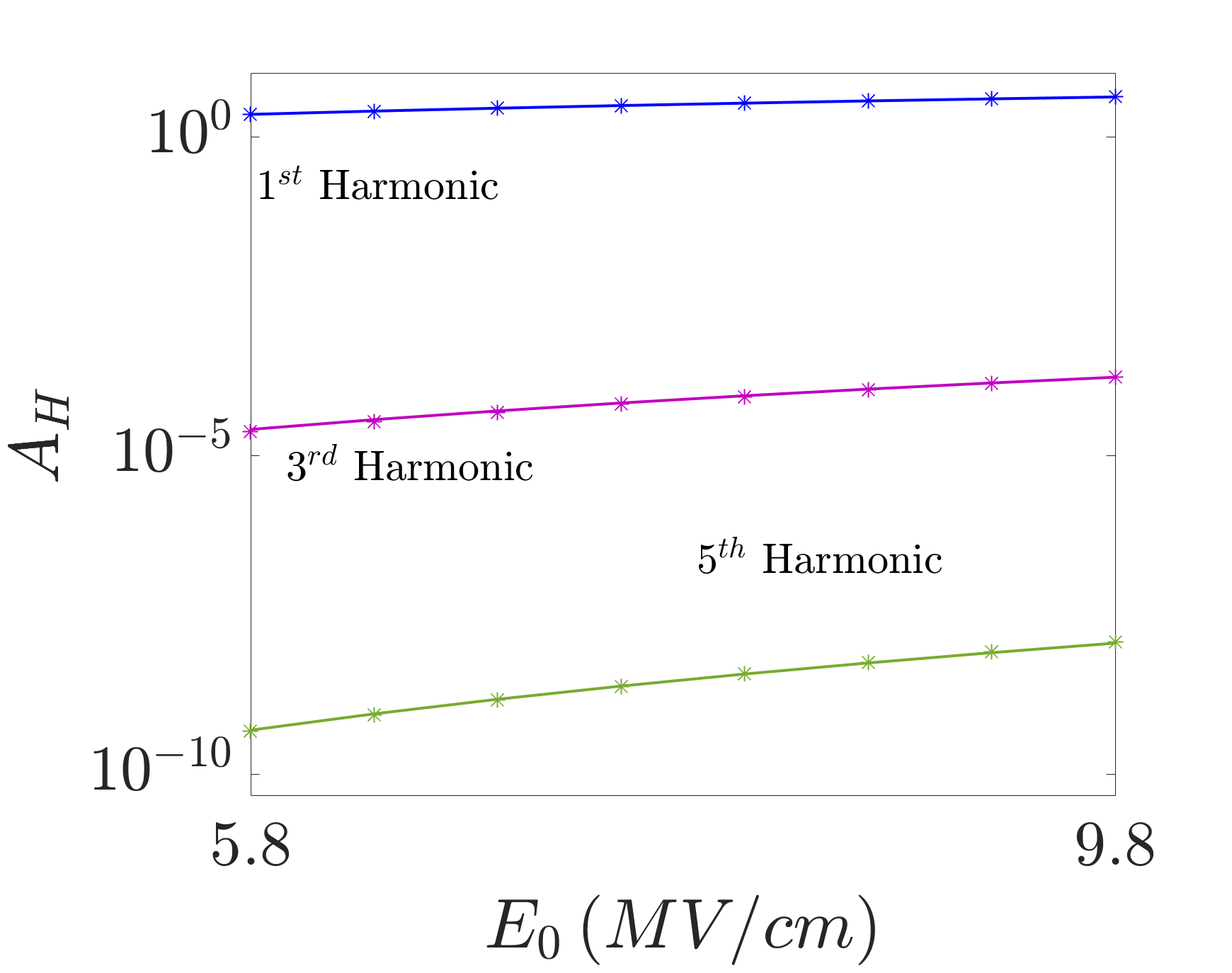}
    \put(0,0){(c)}
    \end{overpic}
     \end{minipage}
    \caption{(a)(b) currents along the $x$ and $y$-directions for different values of the impinging field in logarithmic scale. (c) Variation of harmonic amplitude for the $x$-polarisation with the electric field strength, the solid lines show the theoretical polynomial curves $A_H \approx E_0^n$ in log scale}
    \label{fig2}
\end{figure*}

 We first characterize the nonlinear interaction of TBG with an impinging electromagnetic field for different intensities. The external electromagnetic potential is of the form 
\beq
\AAA(t) = \left(\begin{array}{cc} &\frac{A_0}{\omega_0} e^
{-(t/t_0)^2}\sin(\omega_0 t)  \\ \ &\epsilon \frac{A_0}{\omega_0} e^
{-(t/t_0)^2}\sin(\omega_0 t-\eta) \end{array}\right),
\eeq
where $A_0$ is the amplitude of the field and $t_0$ the pulse duration. The parameter $\epsilon$ and the phase $\eta$ control the field polarization, for $\epsilon,\eta=0$ the field is linearly polarized along the $x$-direction, for $\epsilon=1$, $\eta=\pi/2$ is circularly polarized while for an arbitrary value of $\eta$ the polarization is elliptical. We consider the case of gapless flat bands namely $\theta=1.05$ for a linearly polarized incident electric field along the $x$-direction (figure \ref{fig2}a-b). The current spectra behave accordingly to the symmetries of the system. Since the latter is inversion symmetric, the current spectra show only odd harmonics. Figure \ref{fig2}c shows the harmonic amplitude which is defined \cite{ike}
\beq
A^{(n,\mu)}_H=\sum_{\Omega=n-1/2}^{n+1/2} P^\mu(\Omega),
\eeq
where $\Omega=\omega/\omega_0$, $n$ is the harmonic order and $P^\mu$ are the components of the electric polarisation vector which, in time domain, is defined as 
\beq
\bm P(t) = \sum_\kk \bm \mu_\kk(t)\, p^*_\kk(t) + \text{c.c.}
\eeq
The nonlinear response is in line with perturbation theory $A_H \approx E_0^n$. It is useful at this point to compare the result obtained so far with previous theoretical studies, in particular with ref. \onlinecite{ike}. In \cite{ike} the high harmonic response is studied in a specific commensurate configuration ($\theta=21.79^\circ$) by solving directly the time dependent Schr\"odinger equation. An interesting result is the emergence of dynamical symmetries coupled with the standard symmetries of the lattice ($C_{2y}$ and $C_3$). This generates characteristic selection rules for which even (odd) harmonics are permitted (forbidden) in the $J_x$ ($J_y$) current. This selection rule are not present here because of the additional symmetry constraints that emerge in the low angle regime \cite{zp,am}. In particular the valley degeneracy around the $K$ points \cite{am}.

\begin{figure*}[!t]
        
        \begin{minipage}[b]{.32 \textwidth}
         \centering
         \begin{overpic}[width=\linewidth]{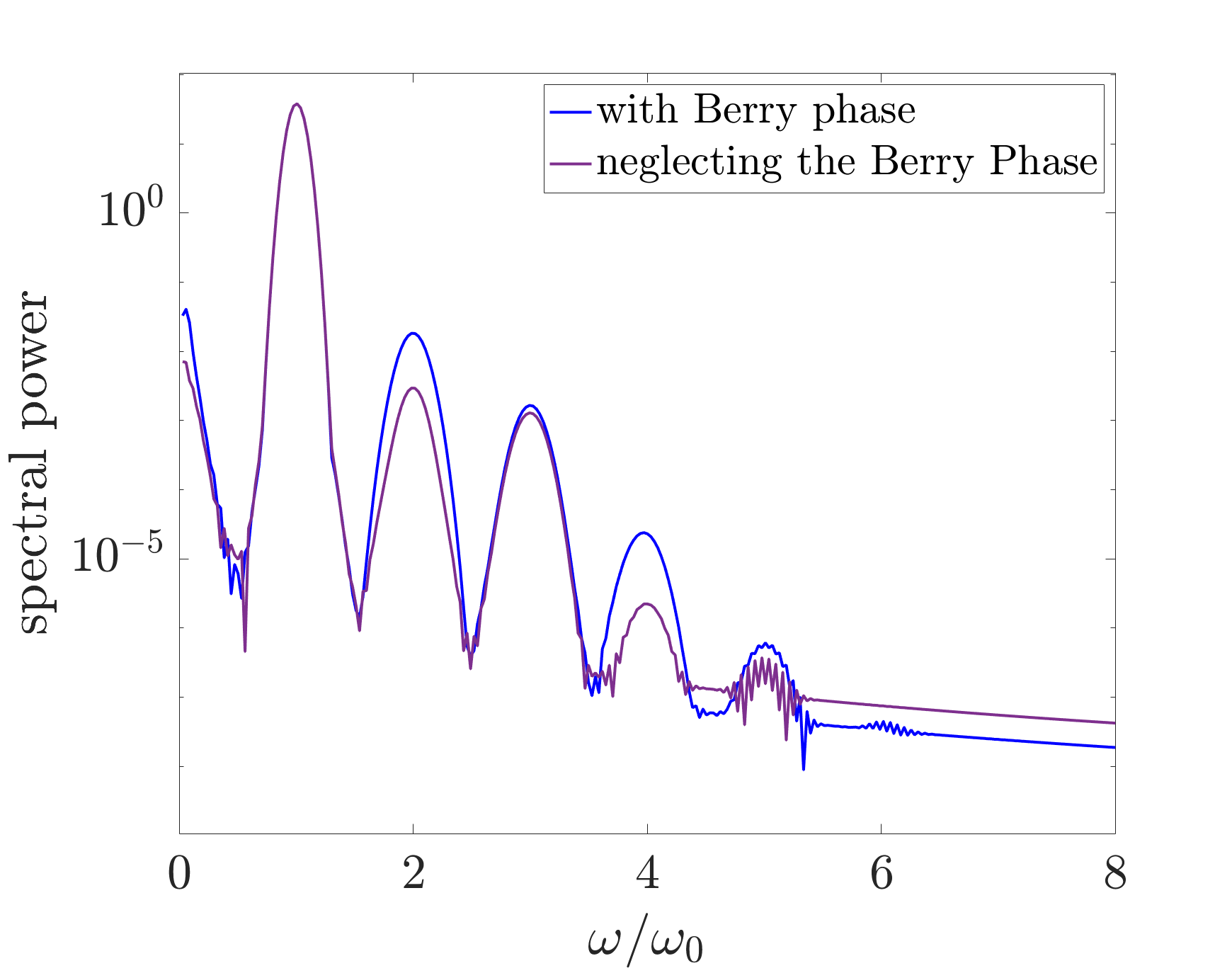}
         \put(0,0){(a)}
    \end{overpic}
     \end{minipage}\hfill
     \begin{minipage}[b]{ .32\textwidth}
         \centering
         \begin{overpic}[width=\linewidth]{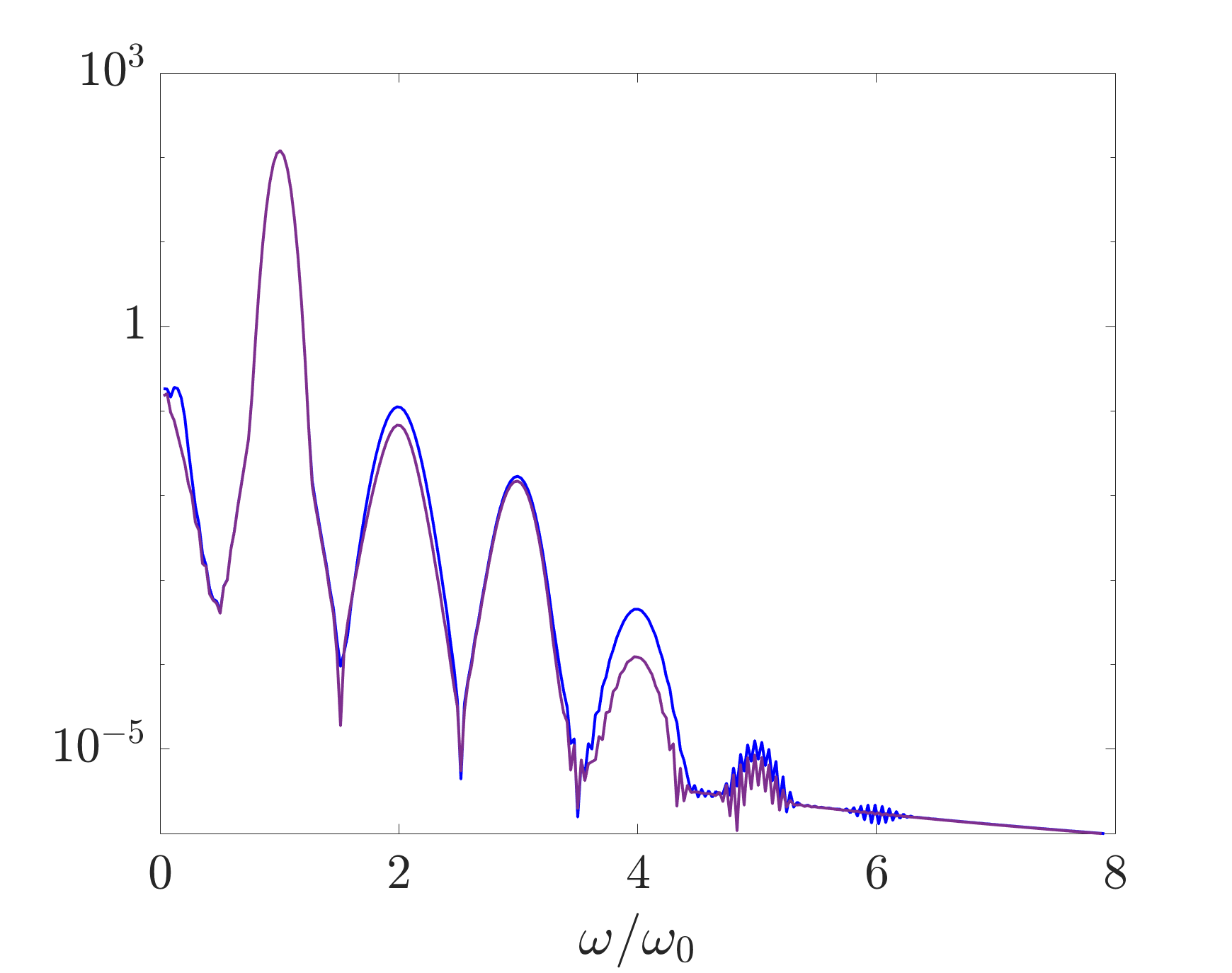}
         \put(0,0){(b)}
    \end{overpic}
     \end{minipage}\hfill
     \begin{minipage}[b]{ .32\textwidth}
         \centering
    \begin{overpic}[width=1.0\linewidth]{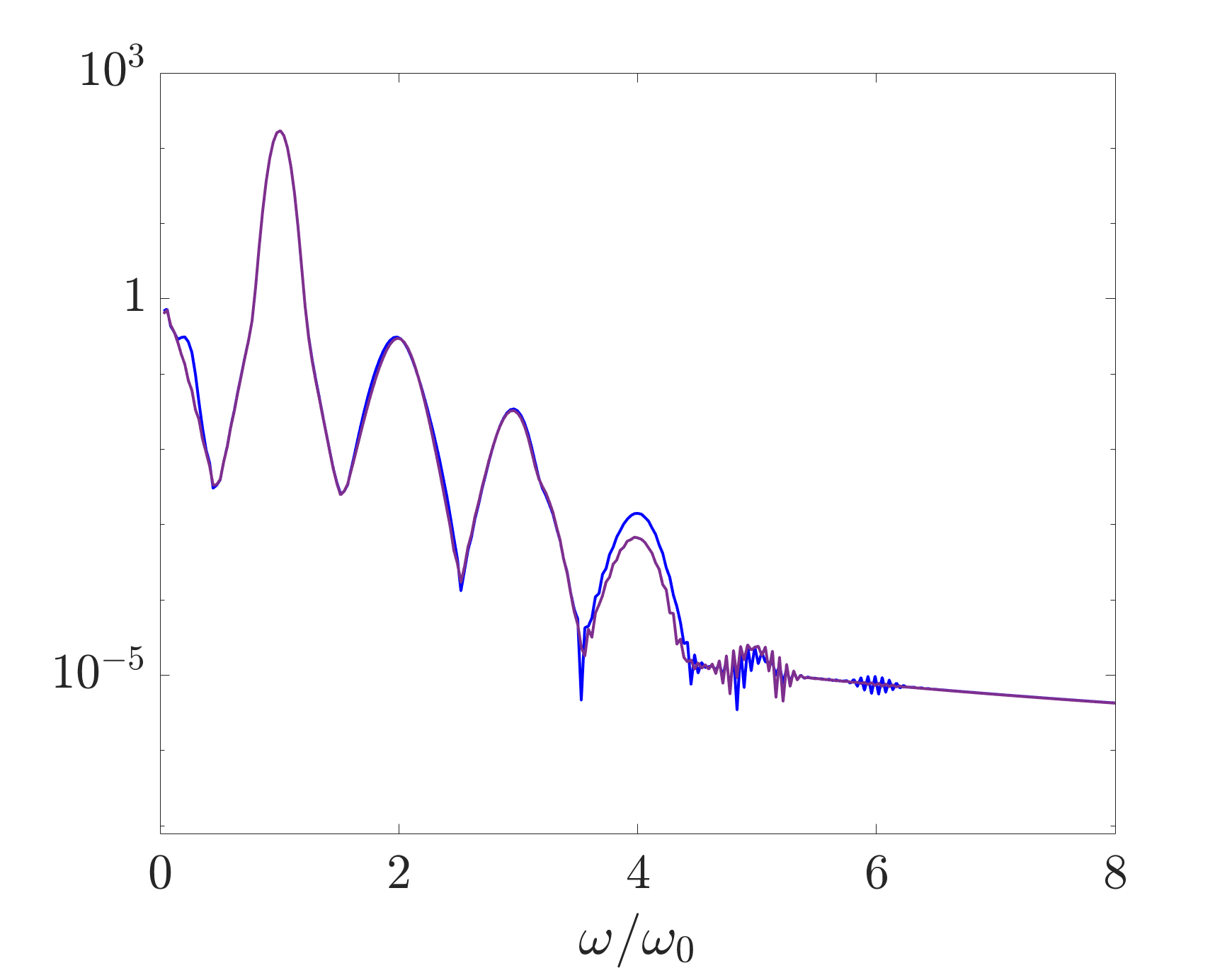}
    \put(0,0){(c)}
    \end{overpic}
     \end{minipage}
        \caption{ Current spectra for (a) $\theta=1.35^\circ$ (b) $\theta=1.85^\circ$ and (c) $\theta=2.15^\circ$ (bottom) with and without the Berry phase} 
        \label{fig4}
\end{figure*}

In what follows we study what happens when we open an inversion-symmetry-breaking energy gap. When a gap is opened in each of the two monolayers, the same happens in the moir\`e band structure. This effectively breaks the inversion symmetry of the system, due to the inequivalence of the two valleys, triggering the presence of even harmonics. These are forbidden in inversion symmetric system due to selection rules in the leading electric dipole contribution \cite{BOYD20201}. An interesting aspect of the gapped case is that it elucidates the role of the Berry phase in the nonlinear response. In figure \ref{fig4} we show the current spectra with and without the Berry phase. We  notice that the Berry phase enhances considerably the even order non-linearity. This has to be expected. The role of the  Berry phase in the nonlinear dynamics is related to the valley inequivalence \cite{cm}, as in layman terms the latter can be considered a measure of the inversion symmetry braking. In the low energy continuum limit  the inversion  symmetry is represented by the simultaneous exchange of valley and sublattice indices \cite{mck}. For this reason we can expect even harmonics to be significantly dependent on the Berry phase terms in the current. At the same time we can see that this effect depends on the geometry in real space, {\it i.e.} it is stronger for smaller angles. 

\begin{figure}[htbp]
    \includegraphics[width=1 \columnwidth]{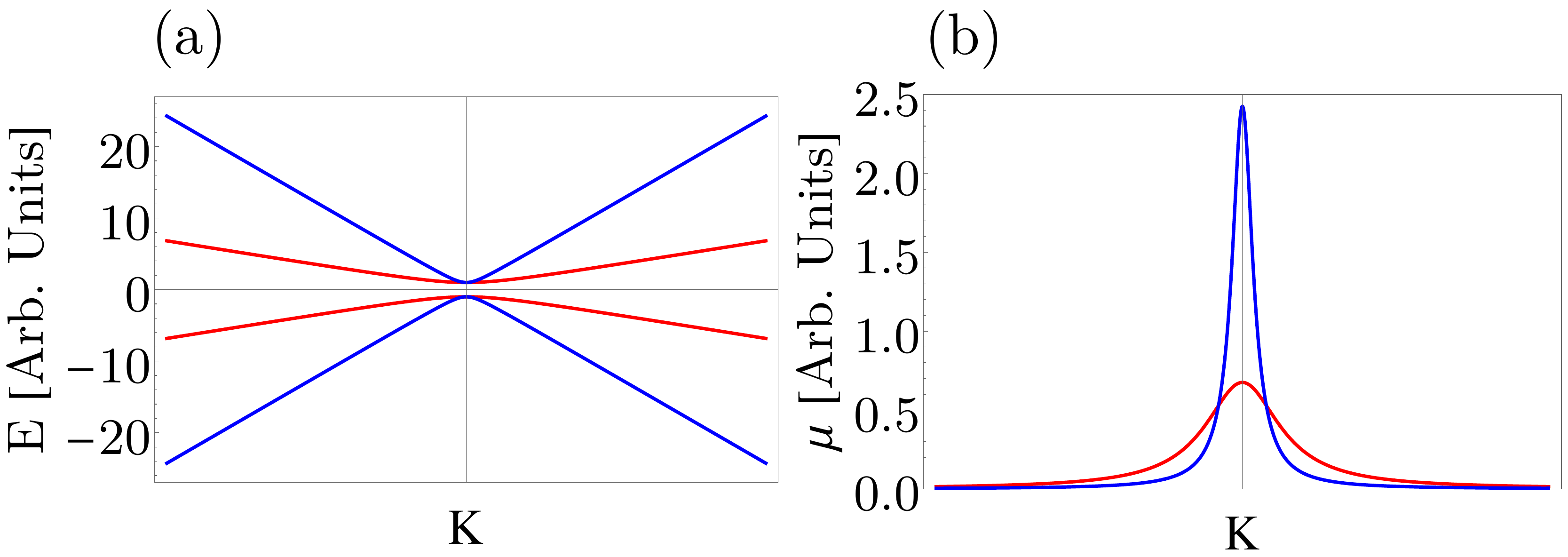}
    \caption{Spectrum and dipole moments for the two lowest energy bands at different angles in the maddive-Dirac-fermion approximation. The red curve is for $\theta=1.35^\circ$, while the blue one is for $\theta=2.35^\circ$.}
    \label{fig3}
\end{figure}

This effect is due to the fact that the slope of the massive-Dirac-fermion energy dispersion increases with twist angle, thus causing a sharper decay of the dipole moment around the $K$-points. 
In fact, around the $K$ point and larger than the magic angle, we can approximate the energy spectrum and dipole moment (along the real-space $x$-direction)
as \cite{cm,vg}
\beq
\begin{aligned}
E^{\lambda}_{\kk}(\theta) &\approx \lambda \sqrt{(\theta\, v_F\, \kk)^2 + (\Delta/2)^2},\\
\mu_{\kk,x}(\theta) &\approx e v_F \Biggl( \frac{\sin \vartheta_\kk}{E_\kk(\theta)} +i\Delta \frac{\cos \vartheta_\kk}{E^2_{\kk}(\theta)} \Biggl),
\end{aligned}
\eeq
where $\vartheta_\kk=\arctan(k_y/k_x)$ is the polar angle and $\Delta$ is the energy gap of the moir\`e band structure. In figure \ref{fig3} we show a qualitative plot of these quantities for $k_y=0$ and two values of the twist angle. We  see that, as the twist angle increases, the massive-Dirac-fermion energy dispersion becomes steeper. This in turn translates into a sharper decay of the dipole $\mu_{\kk,x}(\theta)$.
This is because, at larger angles, less states contribute to the interband current. 

\subsection{Extreme nonlinear optics}
We now focus on the extreme nonlinear optics regime, which means probing the system with an ultrashort pulse with relatively high intensity. In the following, we show numerical results obtained by taking a pulse duration of $5 \, {\rm fs}$ and a pulse intensity of $284 \, {\rm GW/cm^2}$. Under these conditions, the system is in the extreme nonlinear optics regime. In this regime the perturbative expansion of the polarisation in terms of the electric fields fails and new effects emerge \cite{mt,tmw}. To explore this scenario it is worthwhile to study the case of gapped flat bands, a situation in which the system shares some properties with a pure collection of two level systems. In this case, in particular due to the flatness of the bands, the valleys non-equivalence is greatly reduced and inversion symmetry is effectively recovered, at least within the $\kk \cdot \pp$ (low energy) approximation employed here.  In regard to this, the most peculiar effect is the so called odd harmonics \emph{in disguise} of even harmonics, which have been theoretically described \cite{mt} and experimentally observed in  thin ZnO films \cite{tmw}. It is a phenomenon typical of the non-perturbative regime. 

While the even-order susceptibilities are always bound to vanish because of inversion symmetry~\cite{BOYD20201}, in extreme non-linear optics this does not necessarily imply that peaks at even frequencies cannot be generated.
In a certain sense, in this regime the constraints of inversion symmetry, which are quite strong in traditional ({\it i.e.} perturbative) nonlinear optics, are relaxed.
In fact, in traditional nonlinear optics, the spectral width of higher harmonics is much smaller than the carrier frequency $\omega_0$. For this reason there is no interference effect that could generate a peak at even spectral frequency. On the contrary, in extreme nonlinear optics the spectral width of higher harmonics is much broader and can approach $\omega_0$. Thus, odd harmonics envelopes can generate lower harmonics sidebands if they are resonant with transitions frequencies between electronic energy bands.

\begin{figure}[htbp]
    \includegraphics[width=1 \columnwidth]{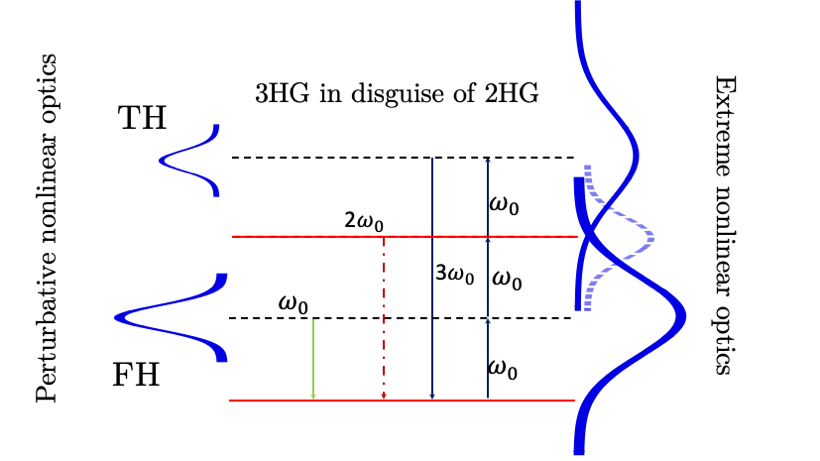}
    \caption{Pictorial representation of third harmonics generation and third harmonics in disguise of second harmonics in a two level system. The red lines represent the electron states, the black dashed lines are the virtual states where the nonlinear frequency mixing takes place. In both cases we assume that the transition frequency is on resonance with twice the carrier frequency. In the case of standard THG the waves are well separated due to the low spectral width with respect to the carrier frequency. In extreme nonlinear optics peaks are much more broad and can interfere generating a peak at twice the carrier frequency.}
    \label{fig5}
\end{figure}

\begin{figure}[htbp]
    \begin{minipage}[b]{.50\linewidth}
         \centering
         \begin{overpic}[width=\textwidth]{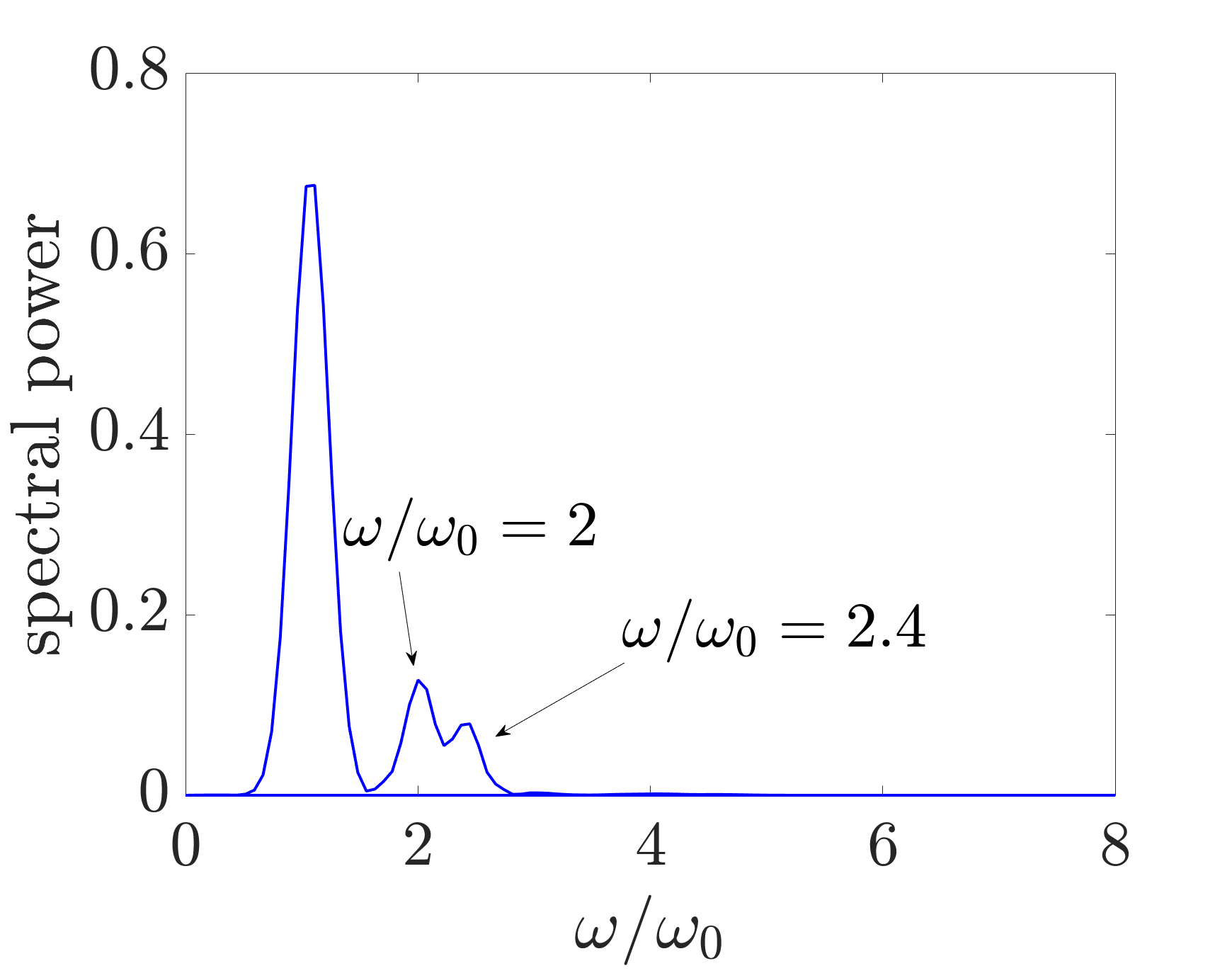}
         \put(0,0){(a)}
    \end{overpic}
     \end{minipage}\hfill
     \begin{minipage}[b]{ 0.50\linewidth}
         \centering
         \begin{overpic}[width=\textwidth]{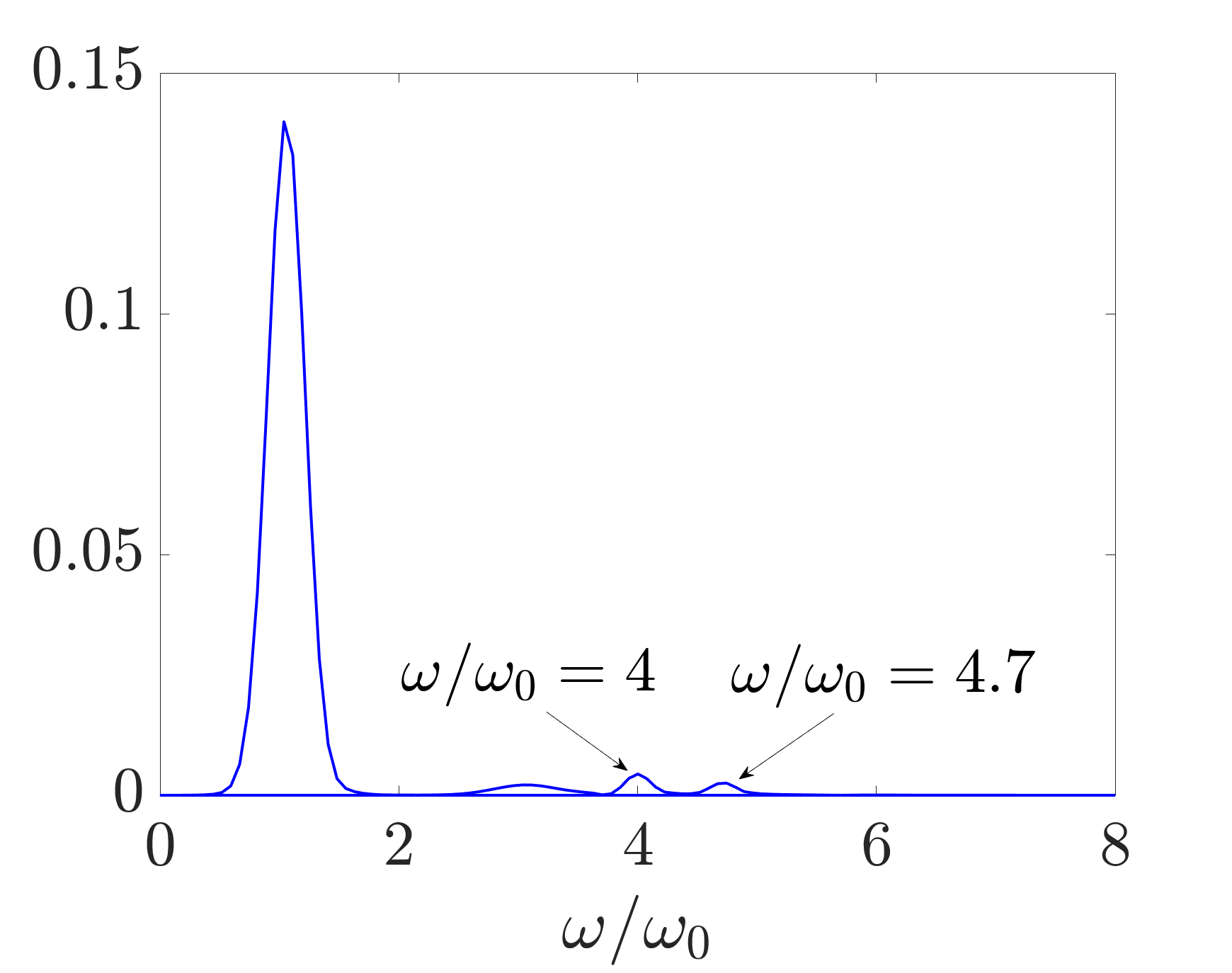}
         \put(0,0){(b)}
    \end{overpic}
     \end{minipage}\hfill
    \caption{Current spectra for a four flat bands system with an impinging field with pulse duration $t_0=5\,{\rm fs}$ and $I_0=284\, {\rm GW/cm}^2$ (a) Third Harmonics in disguise of second harmonics. (b) Fifth harmonics in disguise of fourth harmonics.}
    \label{fig6}
\end{figure}

The effect is pictorially shown in figure \ref{fig5}. We show a two level system with a transition frequency resonating with twice the carrier frequency $\omega_0$.
The second harmonic appears when the first and third harmonic peaks are broad enough that they can interfere, generating a peak at frequency $2\omega_0$. 
If the laser pulse is short, the high-energy tail of the fundamental-harmonic peak and the low-energy tail of the third-harmonic peak meet at around twice the laser center frequency (see figure \ref{fig5}b). 
As the transition frequency between energy bands is twice as large as the laser central frequency, a peak appears at the frequency $\omega = 2 \omega_0$ \cite{tmw}.
This phenomenon is called odd harmonics in disguise of even harmonics \cite{tmw}. We report this effect for the case of three flat bands with dimensionless transition frequencies $\Delta_1/\omega_0 = 2$, $\Delta_2/\omega_0=2.4$ and $\Delta_1/\omega_0 = 4$, $\Delta_2/\omega_0=4.7$. Here $\Delta_{1,2}$ represent the energy gaps between the bands. The carrier frequency in our simulations is tuned so that the field interacts with the three lowest flat bands only.  The energy gap between the two lowest bands is $\Delta_1=0.015\, {\rm eV}$. In figure~\ref{fig6} we can observe the emergence of third and fourth harmonic in disguise, the peak splitting is due to the presence of two transition frequencies resonating with the impinging field. Clearly the fourth harmonic peak is considerably lower as it scales with the fifth order nonlinear susceptibilty $\chi^{(5)}$ rather then $\chi^{(3)}$ \cite{BOYD20201}.

\begin{figure}[htbp!]
    \includegraphics[width=1 \columnwidth]{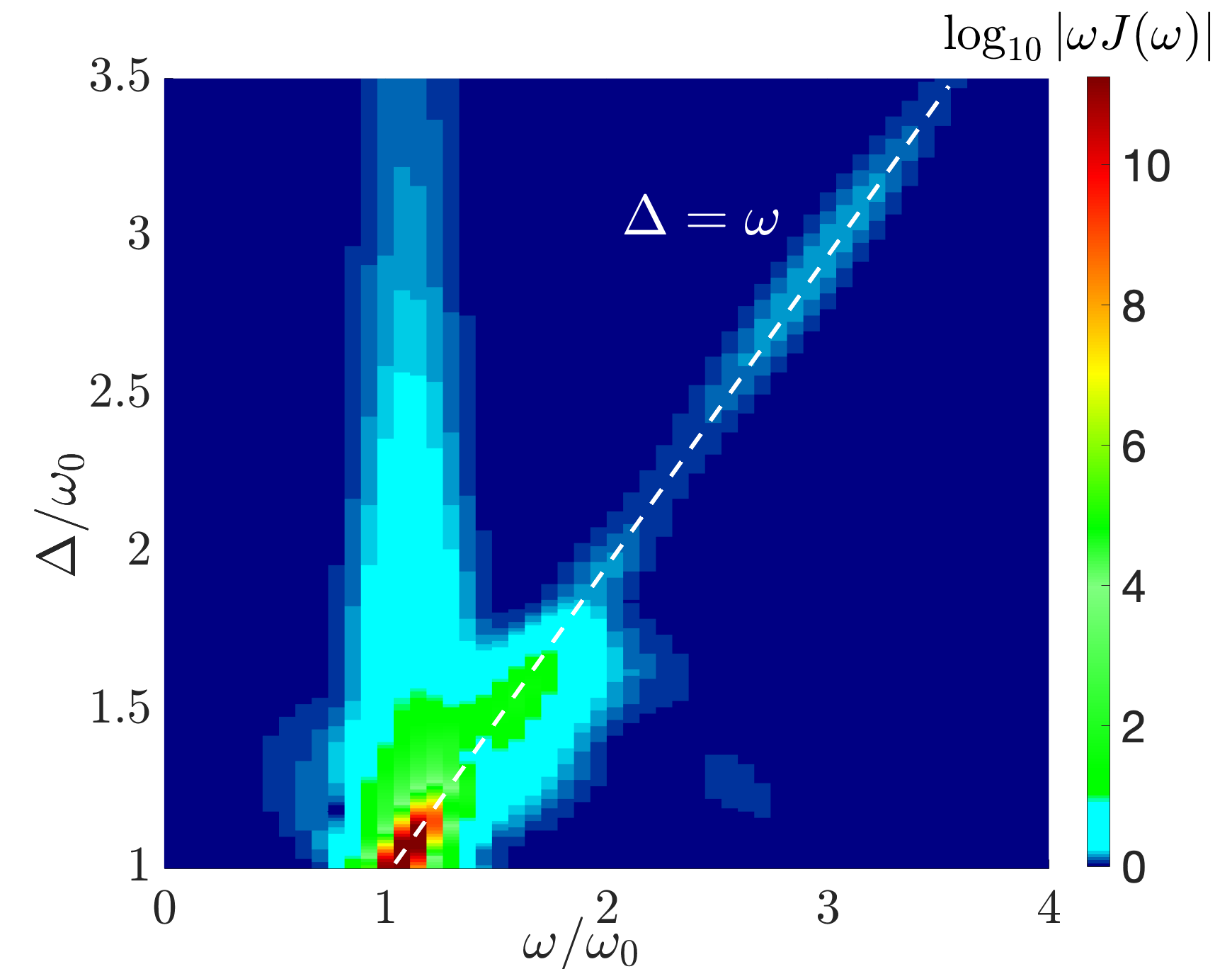}
    \caption{Current spectra for a two flat bands system with an impinging field with pulse duration $t_0=5\,{\rm fs}$ and $I_0=284\, {\rm GW/cm}^2$ and varying transition frequency.}
    \label{fig7}
\end{figure}

To clarify how harmonics in disguise behave while varying the system energy gap (transition frequency), In figure \ref{fig7} we considered two flat bands, with a gap ranging from $\Delta=\omega_0$ to $\Delta=3.5 \omega_0$ interacting with an impinging  laser frequency  $\omega_0=0.015\, {\rm eV}$.
The white dashed line is the resonance line $\omega=\Delta$, where we expect to observe the harmonics in disguise. The strongest peak, as foreseeable, is obtained when the laser frequency is resonant with the band gap, {\it i.e.} $\Delta=\omega_0$. Higher-order harmonics scale with nonlinear susceptibilities which is considerably smaller than the linear one \cite{BOYD20201}. The response around the second and third harmonics, when on resonance, is similar in magnitude which implies that they are both third order effects. This is a strong indication that the second harmonic signal cannot be related to symmetry properties and is indeed a higher harmonic in disguise. 

Another way to confirm this is the case is to compare the
second harmonic signal in figure~\ref{fig6} with the one in figure~\ref{fig4}. In the latter, we see
that the second harmonic is always paired with a zeroth order peak, because sum frequency generation  ($\omega_0+\omega_0$) and difference frequency generation ($\omega_0-\omega_0$) occur with the same probability. On the other hand, in figure~\ref{fig6} and~\ref{fig7} the zeroth order peak is absent, meaning that there is no second-order sum frequency generation process involved in the appearance of a second-harmonic peak.

\section{Conclusion}

 We characterized the nonlinear optical response of TBG in the framework of the Dirac-Bloch equation which is a new method in the context of twisted materials. We elucidated the contribution to the current spectra of the Berry phase and its relation to intraband and interband transitions when inversion symmetry is explicitly broken. The observed effect shines further light on the complex interplay between the lattice geometry in real space, the eigenstate geometry in momentum space and the optical response, which has recently attracted significant attention. Lastly we focused on a phenomenon which is purely non-perturbative and peculiar of extreme nonlinear optics; the generation of harmonics in disguise. We have shown that this can be efficiently realised in a TBLG sample at the \emph{magic angle}, i.e. when the low energy bands of the system are flat. These findings provide further evidence that TBLG is an interesting platform for nonlinear optics in which the response is highly tunable due to the close relation between the twisting and the strength of the dipole coupling. The method and the formalism developed in this paper is flexible enough to be applied to a variety of electronic and magnetic systems.
 
\section*{Acknowledgements}
We acknowledge support from the European Commission under the EU Horizon 2020 MSCA-RISE-2019 programme (project 873028 HYDROTRONICS) and of the Leverhulme Trust under the grant RPG-2019-363.
\appendix*
\section{Generalised Dirac-Bloch equation for an arbitrary number of bands}
The starting point is the Dirac equation for the instantaneous plan-wave expansion of the TBG Hamiltonian, equation (\ref{Dirac}) of the main text, that we report here together with the expansion in the base of the instantaneous eigenstates 
\beq \label{App1}
i\partial_t \psi_{\kk,\qq_i}(t) = \sum_{\qq_j} [(h^{(0)}_{\kk+e\AAA(t),\qq_i,\qq_j}+ V^{tw}_0)\delta_{\qq_i,\qq_j} + V^{tw}_{\qq_i-\qq_j}] \psi_{\kk,\qq_j}(t),
\eeq
\beq \label{App2}
\psi_{\kk,\qq_i}(t) = \sum_{\lambda} c^\lambda_{\kk}(t) \varphi^\lambda_{\kk,\qq_i}(t) e^{-i(\int_{-\infty}^t dt'\,\epsilon^\lambda_\kk(t')+\gamma^\lambda_{\kk}(t))},
\eeq
substituting equation (\ref{App2}) into (\ref{App1}) one gets 
\beq \label{App3}
\dot c^\lambda_\kk(t)=-\sum_{\lambda'=1}^{4N}\sum_{\qq_i} \varphi^{\lambda',\dagger}_{\kk,\qq_i}(t)\cdot \dot {\varphi}^{\lambda}_{\kk,\qq_i}(t) e^{i(\delta \gamma_\kk(t) - \delta E_\kk(t))},
\eeq
where we have defined 
\beq
\begin{aligned}
\delta \gamma_{\kk}(t) &= \gamma^{\lambda'}_\kk(t) - \gamma^\lambda_\kk(t) \\
\delta E_\kk(t)&=\int_{-\infty}^t dt' \, (\epsilon_\kk^{\lambda'}(t') - \epsilon_\kk^\lambda(t'))
\end{aligned}
\eeq
and $N$ is the number of plane-waves. Using the definition of population inversion and microscopic polarisation given in equation (\ref{var}) together with equation (\ref{App3}), we can derive the generalised DBEs
\begin{widetext}
\beq \label{RDB}
\begin{cases}
\dot p^{\lambda,\lambda'}_{\kk} &= -i[\omega_0-\delta\epsilon_{\kk}(t)] p_{\kk} - i \Omega^{\lambda,\lambda'}_{\kk}(t)\,e^{-i\delta\gamma^{ \lambda,\lambda'}_{\kk}+i\omega_0 t} w_\kk -\sum_{\bar \lambda \neq \lambda'} \Omega^{\lambda,\bar \lambda}_{\kk}(t)p_\kk^{\lambda'\bar \lambda}, \\
\\
\dot w^{\lambda,\lambda'}_{\kk} &= - 2\sum_{\bar \lambda}\operatorname{Re}\left\{\left(\Omega^{\bar \lambda,\lambda}_{\kk}(t)\right)^*\,e^{i\delta\gamma^{\bar \lambda,\lambda}_{\kk}+i\omega_0t}p^{\bar \lambda,\lambda}_{\kk}+\left(\Omega^{\bar \lambda,\lambda'}_{\kk}(t)\right)^*\,e^{i\delta\gamma^{\bar \lambda,\lambda'}_{\kk}+i\omega_0t}p^{\bar \lambda,\lambda'}_{\kk}\right\}.
\end{cases}
\eeq
\end{widetext}
This is a $4N(4N-1)$ system of differential equations which accounts for all possible bands couplings. In general it would be very hard to solve even for a relatively small number of plane-waves, due to the high number of coupled bands involved. Assumptions on the physics of the system can help reducing the size of the problem and hence the computational cost. Commonly, one can ignore the dynamics of occupied states below the Fermi energy and consider external fields with frequencies resonating with a limited number of states.

\begin{thebibliography}{36}%
\makeatletter
\providecommand \@ifxundefined [1]{%
 \@ifx{#1\undefined}
}%
\providecommand \@ifnum [1]{%
 \ifnum #1\expandafter \@firstoftwo
 \else \expandafter \@secondoftwo
 \fi
}%
\providecommand \@ifx [1]{%
 \ifx #1\expandafter \@firstoftwo
 \else \expandafter \@secondoftwo
 \fi
}%
\providecommand \natexlab [1]{#1}%
\providecommand \enquote  [1]{``#1''}%
\providecommand \bibnamefont  [1]{#1}%
\providecommand \bibfnamefont [1]{#1}%
\providecommand \citenamefont [1]{#1}%
\providecommand \href@noop [0]{\@secondoftwo}%
\providecommand \href [0]{\begingroup \@sanitize@url \@href}%
\providecommand \@href[1]{\@@startlink{#1}\@@href}%
\providecommand \@@href[1]{\endgroup#1\@@endlink}%
\providecommand \@sanitize@url [0]{\catcode `\\12\catcode `\$12\catcode
  `\&12\catcode `\#12\catcode `\^12\catcode `\_12\catcode `\%12\relax}%
\providecommand \@@startlink[1]{}%
\providecommand \@@endlink[0]{}%
\providecommand \url  [0]{\begingroup\@sanitize@url \@url }%
\providecommand \@url [1]{\endgroup\@href {#1}{\urlprefix }}%
\providecommand \urlprefix  [0]{URL }%
\providecommand \Eprint [0]{\href }%
\providecommand \doibase [0]{https://doi.org/}%
\providecommand \selectlanguage [0]{\@gobble}%
\providecommand \bibinfo  [0]{\@secondoftwo}%
\providecommand \bibfield  [0]{\@secondoftwo}%
\providecommand \translation [1]{[#1]}%
\providecommand \BibitemOpen [0]{}%
\providecommand \bibitemStop [0]{}%
\providecommand \bibitemNoStop [0]{.\EOS\space}%
\providecommand \EOS [0]{\spacefactor3000\relax}%
\providecommand \BibitemShut  [1]{\csname bibitem#1\endcsname}%
\let\auto@bib@innerbib\@empty
\bibitem [{\citenamefont {Nimbalkar}\ and\ \citenamefont {Kim}(2020)}]{nim}%
  \BibitemOpen
  \bibfield  {author} {\bibinfo {author} {\bibfnamefont {A.}~\bibnamefont
  {Nimbalkar}}\ and\ \bibinfo {author} {\bibfnamefont {H.}~\bibnamefont
  {Kim}},\ }\bibfield  {title} {\bibinfo {title} {Opportunities and challenges
  in twisted bilayer graphene: A review},\ }\href
  {https://doi.org/10.1007/s40820-020-00464-8} {\bibfield  {journal} {\bibinfo
  {journal} {Nano-Micro Letters}\ }\textbf {\bibinfo {volume} {12}},\ \bibinfo
  {pages} {126} (\bibinfo {year} {2020})}\BibitemShut {NoStop}%
\bibitem [{\citenamefont {Cao}\ \emph {et~al.}(2018)\citenamefont {Cao},
  \citenamefont {Fatemi}, \citenamefont {Fang}, \citenamefont {Watanabe},
  \citenamefont {Taniguchi}, \citenamefont {Kaxiras},\ and\ \citenamefont
  {Jarillo-Herrero}}]{cao}%
  \BibitemOpen
  \bibfield  {author} {\bibinfo {author} {\bibfnamefont {Y.}~\bibnamefont
  {Cao}}, \bibinfo {author} {\bibfnamefont {V.}~\bibnamefont {Fatemi}},
  \bibinfo {author} {\bibfnamefont {S.}~\bibnamefont {Fang}}, \bibinfo {author}
  {\bibfnamefont {K.}~\bibnamefont {Watanabe}}, \bibinfo {author}
  {\bibfnamefont {T.}~\bibnamefont {Taniguchi}}, \bibinfo {author}
  {\bibfnamefont {E.}~\bibnamefont {Kaxiras}},\ and\ \bibinfo {author}
  {\bibfnamefont {P.}~\bibnamefont {Jarillo-Herrero}},\ }\bibfield  {title}
  {\bibinfo {title} {Unconventional superconductivity in magic-angle graphene
  superlattices},\ }\href {https://doi.org/10.1038/nature26160} {\bibfield
  {journal} {\bibinfo  {journal} {Nature}\ }\textbf {\bibinfo {volume} {556}},\
  \bibinfo {pages} {43} (\bibinfo {year} {2018})}\BibitemShut {NoStop}%
\bibitem [{\citenamefont {Xie}\ \emph {et~al.}(2019)\citenamefont {Xie},
  \citenamefont {Lian}, \citenamefont {J{\"a}ck}, \citenamefont {Liu},
  \citenamefont {Chiu}, \citenamefont {Watanabe}, \citenamefont {Taniguchi},
  \citenamefont {Bernevig},\ and\ \citenamefont {Yazdani}}]{xie}%
  \BibitemOpen
  \bibfield  {author} {\bibinfo {author} {\bibfnamefont {Y.}~\bibnamefont
  {Xie}}, \bibinfo {author} {\bibfnamefont {B.}~\bibnamefont {Lian}}, \bibinfo
  {author} {\bibfnamefont {B.}~\bibnamefont {J{\"a}ck}}, \bibinfo {author}
  {\bibfnamefont {X.}~\bibnamefont {Liu}}, \bibinfo {author} {\bibfnamefont
  {C.-L.}\ \bibnamefont {Chiu}}, \bibinfo {author} {\bibfnamefont
  {K.}~\bibnamefont {Watanabe}}, \bibinfo {author} {\bibfnamefont
  {T.}~\bibnamefont {Taniguchi}}, \bibinfo {author} {\bibfnamefont {B.~A.}\
  \bibnamefont {Bernevig}},\ and\ \bibinfo {author} {\bibfnamefont
  {A.}~\bibnamefont {Yazdani}},\ }\bibfield  {title} {\bibinfo {title}
  {Spectroscopic signatures of many-body correlations in magic-angle twisted
  bilayer graphene},\ }\href {https://doi.org/10.1038/s41586-019-1422-x}
  {\bibfield  {journal} {\bibinfo  {journal} {Nature}\ }\textbf {\bibinfo
  {volume} {572}},\ \bibinfo {pages} {101} (\bibinfo {year}
  {2019})}\BibitemShut {NoStop}%
\bibitem [{\citenamefont {Bistritzer}\ and\ \citenamefont
  {MacDonald}(2011)}]{BiMac}%
  \BibitemOpen
  \bibfield  {author} {\bibinfo {author} {\bibfnamefont {R.}~\bibnamefont
  {Bistritzer}}\ and\ \bibinfo {author} {\bibfnamefont {A.~H.}\ \bibnamefont
  {MacDonald}},\ }\bibfield  {title} {\bibinfo {title} {Moir{\'e} bands in
  twisted double-layer graphene},\ }\href
  {https://doi.org/10.1073/pnas.1108174108} {\bibfield  {journal} {\bibinfo
  {journal} {PNAS}\ }\textbf {\bibinfo {volume} {108}},\ \bibinfo {pages}
  {12233} (\bibinfo {year} {2011})}\BibitemShut {NoStop}%
\bibitem [{\citenamefont {Shallcross}\ \emph {et~al.}(2010)\citenamefont
  {Shallcross}, \citenamefont {Sharma}, \citenamefont {Kandelaki},\ and\
  \citenamefont {Pankratov}}]{ssk}%
  \BibitemOpen
  \bibfield  {author} {\bibinfo {author} {\bibfnamefont {S.}~\bibnamefont
  {Shallcross}}, \bibinfo {author} {\bibfnamefont {S.}~\bibnamefont {Sharma}},
  \bibinfo {author} {\bibfnamefont {E.}~\bibnamefont {Kandelaki}},\ and\
  \bibinfo {author} {\bibfnamefont {O.~A.}\ \bibnamefont {Pankratov}},\
  }\bibfield  {title} {\bibinfo {title} {Electronic structure of turbostratic
  graphene},\ }\href {https://doi.org/10.1103/PhysRevB.81.165105} {\bibfield
  {journal} {\bibinfo  {journal} {Phys. Rev. B}\ }\textbf {\bibinfo {volume}
  {81}},\ \bibinfo {pages} {165105} (\bibinfo {year} {2010})}\BibitemShut
  {NoStop}%
\bibitem [{\citenamefont {Moon}\ and\ \citenamefont {Koshino}(2012)}]{pm}%
  \BibitemOpen
  \bibfield  {author} {\bibinfo {author} {\bibfnamefont {P.}~\bibnamefont
  {Moon}}\ and\ \bibinfo {author} {\bibfnamefont {M.}~\bibnamefont {Koshino}},\
  }\bibfield  {title} {\bibinfo {title} {Energy spectrum and quantum hall
  effect in twisted bilayer graphene},\ }\href
  {https://doi.org/10.1103/PhysRevB.85.195458} {\bibfield  {journal} {\bibinfo
  {journal} {Phys. Rev. B}\ }\textbf {\bibinfo {volume} {85}},\ \bibinfo
  {pages} {195458} (\bibinfo {year} {2012})}\BibitemShut {NoStop}%
\bibitem [{\citenamefont {Po}\ \emph {et~al.}(2018)\citenamefont {Po},
  \citenamefont {Zou}, \citenamefont {Vishwanath},\ and\ \citenamefont
  {Senthil}}]{pla}%
  \BibitemOpen
  \bibfield  {author} {\bibinfo {author} {\bibfnamefont {H.~C.}\ \bibnamefont
  {Po}}, \bibinfo {author} {\bibfnamefont {L.}~\bibnamefont {Zou}}, \bibinfo
  {author} {\bibfnamefont {A.}~\bibnamefont {Vishwanath}},\ and\ \bibinfo
  {author} {\bibfnamefont {T.}~\bibnamefont {Senthil}},\ }\bibfield  {title}
  {\bibinfo {title} {Origin of mott insulating behavior and superconductivity
  in twisted bilayer graphene},\ }\href
  {https://doi.org/10.1103/PhysRevX.8.031089} {\bibfield  {journal} {\bibinfo
  {journal} {Phys. Rev. X}\ }\textbf {\bibinfo {volume} {8}},\ \bibinfo {pages}
  {031089} (\bibinfo {year} {2018})}\BibitemShut {NoStop}%
\bibitem [{\citenamefont {Koshino}\ \emph {et~al.}(2018)\citenamefont
  {Koshino}, \citenamefont {Yuan}, \citenamefont {Koretsune}, \citenamefont
  {Ochi}, \citenamefont {Kuroki},\ and\ \citenamefont {Fu}}]{kmn}%
  \BibitemOpen
  \bibfield  {author} {\bibinfo {author} {\bibfnamefont {M.}~\bibnamefont
  {Koshino}}, \bibinfo {author} {\bibfnamefont {N.~F.~Q.}\ \bibnamefont
  {Yuan}}, \bibinfo {author} {\bibfnamefont {T.}~\bibnamefont {Koretsune}},
  \bibinfo {author} {\bibfnamefont {M.}~\bibnamefont {Ochi}}, \bibinfo {author}
  {\bibfnamefont {K.}~\bibnamefont {Kuroki}},\ and\ \bibinfo {author}
  {\bibfnamefont {L.}~\bibnamefont {Fu}},\ }\bibfield  {title} {\bibinfo
  {title} {Maximally localized wannier orbitals and the extended hubbard model
  for twisted bilayer graphene},\ }\href
  {https://doi.org/10.1103/PhysRevX.8.031087} {\bibfield  {journal} {\bibinfo
  {journal} {Phys. Rev. X}\ }\textbf {\bibinfo {volume} {8}},\ \bibinfo {pages}
  {031087} (\bibinfo {year} {2018})}\BibitemShut {NoStop}%
\bibitem [{\citenamefont {He}\ \emph {et~al.}(2021)\citenamefont {He},
  \citenamefont {Zhou}, \citenamefont {Ye}, \citenamefont {Cho}, \citenamefont
  {Jeong}, \citenamefont {Meng},\ and\ \citenamefont {Wang}}]{He2021}%
  \BibitemOpen
  \bibfield  {author} {\bibinfo {author} {\bibfnamefont {F.}~\bibnamefont
  {He}}, \bibinfo {author} {\bibfnamefont {Y.}~\bibnamefont {Zhou}}, \bibinfo
  {author} {\bibfnamefont {Z.}~\bibnamefont {Ye}}, \bibinfo {author}
  {\bibfnamefont {S.~H.}\ \bibnamefont {Cho}}, \bibinfo {author} {\bibfnamefont
  {J.}~\bibnamefont {Jeong}}, \bibinfo {author} {\bibfnamefont
  {X.}~\bibnamefont {Meng}},\ and\ \bibinfo {author} {\bibfnamefont
  {Y.}~\bibnamefont {Wang}},\ }\bibfield  {title} {\bibinfo {title}
  {{Moir{\'{e}} Patterns in 2D Materials: A Review}},\ }\href
  {https://pubs.acs.org/doi/abs/10.1021/acsnano.0c10435} {\bibfield  {journal}
  {\bibinfo  {journal} {ACS Nano}\ }\textbf {\bibinfo {volume} {15}},\ \bibinfo
  {pages} {5944} (\bibinfo {year} {2021})}\BibitemShut {NoStop}%
\bibitem [{\citenamefont {Andrei}\ \emph {et~al.}(2021)\citenamefont {Andrei},
  \citenamefont {Efetov}, \citenamefont {Jarillo-Herrero}, \citenamefont
  {MacDonald}, \citenamefont {Mak}, \citenamefont {Senthil}, \citenamefont
  {Tutuc}, \citenamefont {Yazdani},\ and\ \citenamefont {Young}}]{Andrei2021}%
  \BibitemOpen
  \bibfield  {author} {\bibinfo {author} {\bibfnamefont {E.~Y.}\ \bibnamefont
  {Andrei}}, \bibinfo {author} {\bibfnamefont {D.~K.}\ \bibnamefont {Efetov}},
  \bibinfo {author} {\bibfnamefont {P.}~\bibnamefont {Jarillo-Herrero}},
  \bibinfo {author} {\bibfnamefont {A.~H.}\ \bibnamefont {MacDonald}}, \bibinfo
  {author} {\bibfnamefont {K.~F.}\ \bibnamefont {Mak}}, \bibinfo {author}
  {\bibfnamefont {T.}~\bibnamefont {Senthil}}, \bibinfo {author} {\bibfnamefont
  {E.}~\bibnamefont {Tutuc}}, \bibinfo {author} {\bibfnamefont
  {A.}~\bibnamefont {Yazdani}},\ and\ \bibinfo {author} {\bibfnamefont {A.~F.}\
  \bibnamefont {Young}},\ }\bibfield  {title} {\bibinfo {title} {{The marvels
  of moir{\'{e}} materials}},\ }\href
  {https://www.nature.com/articles/s41578-021-00284-1} {\bibfield  {journal}
  {\bibinfo  {journal} {Nature Reviews Materials}\ }\textbf {\bibinfo {volume}
  {6}},\ \bibinfo {pages} {201} (\bibinfo {year} {2021})}\BibitemShut {NoStop}%
\bibitem [{\citenamefont {Topp}\ \emph {et~al.}(2019)\citenamefont {Topp},
  \citenamefont {Jotzu}, \citenamefont {McIver}, \citenamefont {Xian},
  \citenamefont {Rubio},\ and\ \citenamefont {Sentef}}]{tj}%
  \BibitemOpen
  \bibfield  {author} {\bibinfo {author} {\bibfnamefont {G.~E.}\ \bibnamefont
  {Topp}}, \bibinfo {author} {\bibfnamefont {G.}~\bibnamefont {Jotzu}},
  \bibinfo {author} {\bibfnamefont {J.~W.}\ \bibnamefont {McIver}}, \bibinfo
  {author} {\bibfnamefont {L.}~\bibnamefont {Xian}}, \bibinfo {author}
  {\bibfnamefont {A.}~\bibnamefont {Rubio}},\ and\ \bibinfo {author}
  {\bibfnamefont {M.~A.}\ \bibnamefont {Sentef}},\ }\bibfield  {title}
  {\bibinfo {title} {Topological floquet engineering of twisted bilayer
  graphene},\ }\href@noop {} {\bibfield  {journal} {\bibinfo  {journal} {Phys.
  Rev. Research}\ }\textbf {\bibinfo {volume} {1}},\ \bibinfo {pages} {023031}
  (\bibinfo {year} {2019})}\BibitemShut {NoStop}%
\bibitem [{\citenamefont {Vogl}\ \emph
  {et~al.}(2020{\natexlab{a}})\citenamefont {Vogl}, \citenamefont
  {Rodriguez-Vega},\ and\ \citenamefont {Fiete}}]{vm}%
  \BibitemOpen
  \bibfield  {author} {\bibinfo {author} {\bibfnamefont {M.}~\bibnamefont
  {Vogl}}, \bibinfo {author} {\bibfnamefont {M.}~\bibnamefont
  {Rodriguez-Vega}},\ and\ \bibinfo {author} {\bibfnamefont {G.~A.}\
  \bibnamefont {Fiete}},\ }\bibfield  {title} {\bibinfo {title} {Floquet
  engineering of interlayer couplings: Tuning the magic angle of twisted
  bilayer graphene at the exit of a waveguide},\ }\href@noop {} {\bibfield
  {journal} {\bibinfo  {journal} {Phys. Rev. B}\ }\textbf {\bibinfo {volume}
  {101}},\ \bibinfo {pages} {241408} (\bibinfo {year}
  {2020}{\natexlab{a}})}\BibitemShut {NoStop}%
\bibitem [{\citenamefont {Vogl}\ \emph
  {et~al.}(2020{\natexlab{b}})\citenamefont {Vogl}, \citenamefont
  {Rodriguez-Vega},\ and\ \citenamefont {Fiete}}]{vm1}%
  \BibitemOpen
  \bibfield  {author} {\bibinfo {author} {\bibfnamefont {M.}~\bibnamefont
  {Vogl}}, \bibinfo {author} {\bibfnamefont {M.}~\bibnamefont
  {Rodriguez-Vega}},\ and\ \bibinfo {author} {\bibfnamefont {G.~A.}\
  \bibnamefont {Fiete}},\ }\bibfield  {title} {\bibinfo {title} {Effective
  floquet hamiltonians for periodically driven twisted bilayer graphene},\
  }\href {https://doi.org/10.1103/PhysRevB.101.235411} {\bibfield  {journal}
  {\bibinfo  {journal} {Phys. Rev. B}\ }\textbf {\bibinfo {volume} {101}},\
  \bibinfo {pages} {235411} (\bibinfo {year} {2020}{\natexlab{b}})}\BibitemShut
  {NoStop}%
\bibitem [{\citenamefont {Otteneder}\ \emph {et~al.}(2020)\citenamefont
  {Otteneder}, \citenamefont {Hubmann}, \citenamefont {Lu}, \citenamefont
  {Kozlov}, \citenamefont {Golub}, \citenamefont {Watanabe}, \citenamefont
  {Taniguchi}, \citenamefont {Efetov},\ and\ \citenamefont {Ganichev}}]{oh}%
  \BibitemOpen
  \bibfield  {author} {\bibinfo {author} {\bibfnamefont {M.}~\bibnamefont
  {Otteneder}}, \bibinfo {author} {\bibfnamefont {S.}~\bibnamefont {Hubmann}},
  \bibinfo {author} {\bibfnamefont {X.}~\bibnamefont {Lu}}, \bibinfo {author}
  {\bibfnamefont {D.~A.}\ \bibnamefont {Kozlov}}, \bibinfo {author}
  {\bibfnamefont {L.~E.}\ \bibnamefont {Golub}}, \bibinfo {author}
  {\bibfnamefont {K.}~\bibnamefont {Watanabe}}, \bibinfo {author}
  {\bibfnamefont {T.}~\bibnamefont {Taniguchi}}, \bibinfo {author}
  {\bibfnamefont {D.~K.}\ \bibnamefont {Efetov}},\ and\ \bibinfo {author}
  {\bibfnamefont {S.~D.}\ \bibnamefont {Ganichev}},\ }\bibfield  {title}
  {\bibinfo {title} {Terahertz photogalvanics in twisted bilayer graphene close
  to the second magic angle},\ }\href@noop {} {\bibfield  {journal} {\bibinfo
  {journal} {Nano Letters}\ }\textbf {\bibinfo {volume} {20}},\ \bibinfo
  {pages} {7152} (\bibinfo {year} {2020})}\BibitemShut {NoStop}%
\bibitem [{\citenamefont {Gao}\ \emph {et~al.}(2020)\citenamefont {Gao},
  \citenamefont {Zhang},\ and\ \citenamefont {Xiao}}]{gy}%
  \BibitemOpen
  \bibfield  {author} {\bibinfo {author} {\bibfnamefont {Y.}~\bibnamefont
  {Gao}}, \bibinfo {author} {\bibfnamefont {Y.}~\bibnamefont {Zhang}},\ and\
  \bibinfo {author} {\bibfnamefont {D.}~\bibnamefont {Xiao}},\ }\bibfield
  {title} {\bibinfo {title} {Tunable layer circular photogalvanic effect in
  twisted bilayers},\ }\href {https://doi.org/10.1103/PhysRevLett.124.077401}
  {\bibfield  {journal} {\bibinfo  {journal} {Phys. Rev. Lett.}\ }\textbf
  {\bibinfo {volume} {124}},\ \bibinfo {pages} {077401} (\bibinfo {year}
  {2020})}\BibitemShut {NoStop}%
\bibitem [{\citenamefont {Ikeda}(2020)}]{ike}%
  \BibitemOpen
  \bibfield  {author} {\bibinfo {author} {\bibfnamefont {T.~N.}\ \bibnamefont
  {Ikeda}},\ }\bibfield  {title} {\bibinfo {title} {High-order nonlinear
  optical response of a twisted bilayer graphene},\ }\href
  {https://doi.org/10.1103/PhysRevResearch.2.032015} {\bibfield  {journal}
  {\bibinfo  {journal} {Phys. Rev. Research}\ }\textbf {\bibinfo {volume}
  {2}},\ \bibinfo {pages} {032015} (\bibinfo {year} {2020})}\BibitemShut
  {NoStop}%
\bibitem [{\citenamefont {Du}\ \emph {et~al.}(2021)\citenamefont {Du},
  \citenamefont {Liu}, \citenamefont {Zeng},\ and\ \citenamefont {Li}}]{mcz}%
  \BibitemOpen
  \bibfield  {author} {\bibinfo {author} {\bibfnamefont {M.}~\bibnamefont
  {Du}}, \bibinfo {author} {\bibfnamefont {C.}~\bibnamefont {Liu}}, \bibinfo
  {author} {\bibfnamefont {Z.}~\bibnamefont {Zeng}},\ and\ \bibinfo {author}
  {\bibfnamefont {R.}~\bibnamefont {Li}},\ }\bibfield  {title} {\bibinfo
  {title} {High-order harmonic generation from twisted bilayer graphene driven
  by a midinfrared laser field},\ }\href
  {https://doi.org/10.1103/PhysRevA.104.033113} {\bibfield  {journal} {\bibinfo
   {journal} {Phys. Rev. A}\ }\textbf {\bibinfo {volume} {104}},\ \bibinfo
  {pages} {033113} (\bibinfo {year} {2021})}\BibitemShut {NoStop}%
\bibitem [{\citenamefont {Zuber}\ and\ \citenamefont {Zhang}(2021)}]{zz}%
  \BibitemOpen
  \bibfield  {author} {\bibinfo {author} {\bibfnamefont {J.~W.}\ \bibnamefont
  {Zuber}}\ and\ \bibinfo {author} {\bibfnamefont {C.}~\bibnamefont {Zhang}},\
  }\bibfield  {title} {\bibinfo {title} {Nonlinear optical response of twisted
  bilayer graphene},\ }\href {https://doi.org/10.1103/PhysRevB.103.245417}
  {\bibfield  {journal} {\bibinfo  {journal} {Phys. Rev. B}\ }\textbf {\bibinfo
  {volume} {103}},\ \bibinfo {pages} {245417} (\bibinfo {year}
  {2021})}\BibitemShut {NoStop}%
\bibitem [{\citenamefont {Ha}\ \emph {et~al.}(2021)\citenamefont {Ha},
  \citenamefont {Park}, \citenamefont {Kim}, \citenamefont {Shin},
  \citenamefont {Choi}, \citenamefont {Park}, \citenamefont {Moon},
  \citenamefont {Chae}, \citenamefont {Jung}, \citenamefont {Lee},
  \citenamefont {Yoo}, \citenamefont {Park}, \citenamefont {Ahn},\ and\
  \citenamefont {Yeom}}]{sn}%
  \BibitemOpen
  \bibfield  {author} {\bibinfo {author} {\bibfnamefont {S.}~\bibnamefont
  {Ha}}, \bibinfo {author} {\bibfnamefont {N.~H.}\ \bibnamefont {Park}},
  \bibinfo {author} {\bibfnamefont {H.}~\bibnamefont {Kim}}, \bibinfo {author}
  {\bibfnamefont {J.}~\bibnamefont {Shin}}, \bibinfo {author} {\bibfnamefont
  {J.}~\bibnamefont {Choi}}, \bibinfo {author} {\bibfnamefont {S.}~\bibnamefont
  {Park}}, \bibinfo {author} {\bibfnamefont {J.-Y.}\ \bibnamefont {Moon}},
  \bibinfo {author} {\bibfnamefont {K.}~\bibnamefont {Chae}}, \bibinfo {author}
  {\bibfnamefont {J.}~\bibnamefont {Jung}}, \bibinfo {author} {\bibfnamefont
  {J.-H.}\ \bibnamefont {Lee}}, \bibinfo {author} {\bibfnamefont
  {Y.}~\bibnamefont {Yoo}}, \bibinfo {author} {\bibfnamefont {J.-Y.}\
  \bibnamefont {Park}}, \bibinfo {author} {\bibfnamefont {K.~J.}\ \bibnamefont
  {Ahn}},\ and\ \bibinfo {author} {\bibfnamefont {D.-I.}\ \bibnamefont
  {Yeom}},\ }\bibfield  {title} {\bibinfo {title} {Enhanced third-harmonic
  generation by manipulating the twist angle of bilayer graphene},\ }\href@noop
  {} {\bibfield  {journal} {\bibinfo  {journal} {Light: Science \&
  Applications}\ }\textbf {\bibinfo {volume} {10}},\ \bibinfo {pages} {19}
  (\bibinfo {year} {2021})}\BibitemShut {NoStop}%
\bibitem [{\citenamefont {Ishikawa}(2010)}]{Ishikawa_2010}%
  \BibitemOpen
  \bibfield  {author} {\bibinfo {author} {\bibfnamefont {K.~L.}\ \bibnamefont
  {Ishikawa}},\ }\bibfield  {title} {\bibinfo {title} {Nonlinear optical
  response of graphene in time domain},\ }\href
  {https://doi.org/10.1103/PhysRevB.82.201402} {\bibfield  {journal} {\bibinfo
  {journal} {Physical Review B}\ }\textbf {\bibinfo {volume} {82}},\ \bibinfo
  {pages} {201402} (\bibinfo {year} {2010})}\BibitemShut {NoStop}%
\bibitem [{\citenamefont {Carvalho}\ \emph {et~al.}(2018)\citenamefont
  {Carvalho}, \citenamefont {Marini},\ and\ \citenamefont {Biancalana}}]{cb}%
  \BibitemOpen
  \bibfield  {author} {\bibinfo {author} {\bibfnamefont {D.~N.}\ \bibnamefont
  {Carvalho}}, \bibinfo {author} {\bibfnamefont {A.}~\bibnamefont {Marini}},\
  and\ \bibinfo {author} {\bibfnamefont {F.}~\bibnamefont {Biancalana}},\
  }\bibfield  {title} {\bibinfo {title} {The nonlinear optical effects of
  opening a gap in graphene},\ }\href@noop {} {\bibfield  {journal} {\bibinfo
  {journal} {Phys. Rev. B.}\ }\textbf {\bibinfo {volume} {97}},\ \bibinfo
  {pages} {195123} (\bibinfo {year} {2018})}\BibitemShut {NoStop}%
\bibitem [{\citenamefont {Carvalho}\ \emph {et~al.}(2017)\citenamefont
  {Carvalho}, \citenamefont {Marini},\ and\ \citenamefont {Biancalana}}]{cm}%
  \BibitemOpen
  \bibfield  {author} {\bibinfo {author} {\bibfnamefont {D.~N.}\ \bibnamefont
  {Carvalho}}, \bibinfo {author} {\bibfnamefont {A.}~\bibnamefont {Marini}},\
  and\ \bibinfo {author} {\bibfnamefont {F.}~\bibnamefont {Biancalana}},\
  }\bibfield  {title} {\bibinfo {title} {Dynamical centrosymmetry breaking - a
  novel mechanism for second harmonic generation in graphene},\ }\href@noop {}
  {\bibfield  {journal} {\bibinfo  {journal} {Annals of Physics}\ }\textbf
  {\bibinfo {volume} {378}},\ \bibinfo {pages} {24} (\bibinfo {year}
  {2017})}\BibitemShut {NoStop}%
\bibitem [{\citenamefont {Villari}\ \emph {et~al.}(2018)\citenamefont
  {Villari}, \citenamefont {Galbraith},\ and\ \citenamefont {Biancalana}}]{vg}%
  \BibitemOpen
  \bibfield  {author} {\bibinfo {author} {\bibfnamefont {L.~D.~M.}\
  \bibnamefont {Villari}}, \bibinfo {author} {\bibfnamefont {I.}~\bibnamefont
  {Galbraith}},\ and\ \bibinfo {author} {\bibfnamefont {F.}~\bibnamefont
  {Biancalana}},\ }\bibfield  {title} {\bibinfo {title} {Coulomb effects in the
  absorbance spectra of two-dimensional \upp{D}irac materials},\ }\href@noop {}
  {\bibfield  {journal} {\bibinfo  {journal} {Phys. Rev. B}\ }\textbf {\bibinfo
  {volume} {98}},\ \bibinfo {pages} {205402} (\bibinfo {year}
  {2018})}\BibitemShut {NoStop}%
\bibitem [{\citenamefont {Mitscherling}(2020)}]{jm}%
  \BibitemOpen
  \bibfield  {author} {\bibinfo {author} {\bibfnamefont {J.}~\bibnamefont
  {Mitscherling}},\ }\bibfield  {title} {\bibinfo {title} {Longitudinal and
  anomalous hall conductivity of a general two-band model},\ }\href
  {https://doi.org/10.1103/PhysRevB.102.165151} {\bibfield  {journal} {\bibinfo
   {journal} {Phys. Rev. B}\ }\textbf {\bibinfo {volume} {102}},\ \bibinfo
  {pages} {165151} (\bibinfo {year} {2020})}\BibitemShut {NoStop}%
\bibitem [{\citenamefont {Mitscherling}\ and\ \citenamefont
  {Holder}(2022)}]{mh}%
  \BibitemOpen
  \bibfield  {author} {\bibinfo {author} {\bibfnamefont {J.}~\bibnamefont
  {Mitscherling}}\ and\ \bibinfo {author} {\bibfnamefont {T.}~\bibnamefont
  {Holder}},\ }\bibfield  {title} {\bibinfo {title} {Bound on resistivity in
  flat-band materials due to the quantum metric},\ }\href
  {https://doi.org/10.1103/PhysRevB.105.085154} {\bibfield  {journal} {\bibinfo
   {journal} {Phys. Rev. B}\ }\textbf {\bibinfo {volume} {105}},\ \bibinfo
  {pages} {085154} (\bibinfo {year} {2022})}\BibitemShut {NoStop}%
\bibitem [{\citenamefont {Lindberg}\ and\ \citenamefont {Koch}(1988)}]{lk}%
  \BibitemOpen
  \bibfield  {author} {\bibinfo {author} {\bibfnamefont {M.}~\bibnamefont
  {Lindberg}}\ and\ \bibinfo {author} {\bibfnamefont {S.~W.}\ \bibnamefont
  {Koch}},\ }\bibfield  {title} {\bibinfo {title} {Effective \uppercase{b}loch
  equations for semiconductors},\ }\href@noop {} {\bibfield  {journal}
  {\bibinfo  {journal} {Phys. Rev. B}\ }\textbf {\bibinfo {volume} {38}},\
  \bibinfo {pages} {3342} (\bibinfo {year} {1988})}\BibitemShut {NoStop}%
\bibitem [{\citenamefont {Tamashevich}\ \emph {et~al.}(2022)\citenamefont
  {Tamashevich}, \citenamefont {Villari},\ and\ \citenamefont
  {Ornigotti}}]{tvo1}%
  \BibitemOpen
  \bibfield  {author} {\bibinfo {author} {\bibfnamefont {Y.}~\bibnamefont
  {Tamashevich}}, \bibinfo {author} {\bibfnamefont {L.~D.~M.}\ \bibnamefont
  {Villari}},\ and\ \bibinfo {author} {\bibfnamefont {M.}~\bibnamefont
  {Ornigotti}},\ }\bibfield  {title} {\bibinfo {title} {Nonlinear optical
  response of type-ii weyl fermions in two dimensions},\ }\href
  {https://doi.org/10.1103/PhysRevB.105.195102} {\bibfield  {journal} {\bibinfo
   {journal} {Phys. Rev. B}\ }\textbf {\bibinfo {volume} {105}},\ \bibinfo
  {pages} {195102} (\bibinfo {year} {2022})}\BibitemShut {NoStop}%
\bibitem [{\citenamefont {Sakurai}\ and\ \citenamefont
  {Napolitano}(2020)}]{sakurai_napolitano_2020}%
  \BibitemOpen
  \bibfield  {author} {\bibinfo {author} {\bibfnamefont {J.~J.}\ \bibnamefont
  {Sakurai}}\ and\ \bibinfo {author} {\bibfnamefont {J.}~\bibnamefont
  {Napolitano}},\ }\href {https://doi.org/10.1017/9781108587280} {\emph
  {\bibinfo {title} {Modern Quantum Mechanics}}},\ \bibinfo {edition} {3rd}\
  ed.\ (\bibinfo  {publisher} {Cambridge University Press},\ \bibinfo {year}
  {2020})\BibitemShut {NoStop}%
\bibitem [{\citenamefont {Ishikawa}(2013)}]{Ishikawa_2013}%
  \BibitemOpen
  \bibfield  {author} {\bibinfo {author} {\bibfnamefont {K.~L.}\ \bibnamefont
  {Ishikawa}},\ }\bibfield  {title} {\bibinfo {title} {Electronic response of
  graphene to an ultrashort intense terahertz radiation pulse},\ }\href
  {https://doi.org/10.1088/1367-2630/15/5/055021} {\bibfield  {journal}
  {\bibinfo  {journal} {New Journal of Physics}\ }\textbf {\bibinfo {volume}
  {15}},\ \bibinfo {pages} {055021} (\bibinfo {year} {2013})}\BibitemShut
  {NoStop}%
\bibitem [{\citenamefont {San-Jose}\ \emph {et~al.}(2012)\citenamefont
  {San-Jose}, \citenamefont {Gonz\'alez},\ and\ \citenamefont {Guinea}}]{gg}%
  \BibitemOpen
  \bibfield  {author} {\bibinfo {author} {\bibfnamefont {P.}~\bibnamefont
  {San-Jose}}, \bibinfo {author} {\bibfnamefont {J.}~\bibnamefont
  {Gonz\'alez}},\ and\ \bibinfo {author} {\bibfnamefont {F.}~\bibnamefont
  {Guinea}},\ }\bibfield  {title} {\bibinfo {title} {Non-abelian gauge
  potentials in graphene bilayers},\ }\href
  {https://doi.org/10.1103/PhysRevLett.108.216802} {\bibfield  {journal}
  {\bibinfo  {journal} {Phys. Rev. Lett.}\ }\textbf {\bibinfo {volume} {108}},\
  \bibinfo {pages} {216802} (\bibinfo {year} {2012})}\BibitemShut {NoStop}%
\bibitem [{\citenamefont {Zou}\ \emph {et~al.}(2018)\citenamefont {Zou},
  \citenamefont {Po}, \citenamefont {Vishwanath},\ and\ \citenamefont
  {Senthil}}]{zp}%
  \BibitemOpen
  \bibfield  {author} {\bibinfo {author} {\bibfnamefont {L.}~\bibnamefont
  {Zou}}, \bibinfo {author} {\bibfnamefont {H.~C.}\ \bibnamefont {Po}},
  \bibinfo {author} {\bibfnamefont {A.}~\bibnamefont {Vishwanath}},\ and\
  \bibinfo {author} {\bibfnamefont {T.}~\bibnamefont {Senthil}},\ }\bibfield
  {title} {\bibinfo {title} {Band structure of twisted bilayer graphene:
  Emergent symmetries, commensurate approximants, and wannier obstructions},\
  }\href {https://link.aps.org/doi/10.1103/PhysRevB.98.085435} {\bibfield
  {journal} {\bibinfo  {journal} {Phys. Rev. B}\ }\textbf {\bibinfo {volume}
  {98}},\ \bibinfo {pages} {085435} (\bibinfo {year} {2018})}\BibitemShut
  {NoStop}%
\bibitem [{\citenamefont {Angeli}\ \emph {et~al.}(2018)\citenamefont {Angeli},
  \citenamefont {Mandelli}, \citenamefont {Valli}, \citenamefont {Amaricci},
  \citenamefont {Capone}, \citenamefont {Tosatti},\ and\ \citenamefont
  {Fabrizio}}]{am}%
  \BibitemOpen
  \bibfield  {author} {\bibinfo {author} {\bibfnamefont {M.}~\bibnamefont
  {Angeli}}, \bibinfo {author} {\bibfnamefont {D.}~\bibnamefont {Mandelli}},
  \bibinfo {author} {\bibfnamefont {A.}~\bibnamefont {Valli}}, \bibinfo
  {author} {\bibfnamefont {A.}~\bibnamefont {Amaricci}}, \bibinfo {author}
  {\bibfnamefont {M.}~\bibnamefont {Capone}}, \bibinfo {author} {\bibfnamefont
  {E.}~\bibnamefont {Tosatti}},\ and\ \bibinfo {author} {\bibfnamefont
  {M.}~\bibnamefont {Fabrizio}},\ }\bibfield  {title} {\bibinfo {title}
  {Emergent ${D}_{6}$ symmetry in fully relaxed magic-angle twisted bilayer
  graphene},\ }\href {https://link.aps.org/doi/10.1103/PhysRevB.98.235137}
  {\bibfield  {journal} {\bibinfo  {journal} {Phys. Rev. B}\ }\textbf {\bibinfo
  {volume} {98}},\ \bibinfo {pages} {235137} (\bibinfo {year}
  {2018})}\BibitemShut {NoStop}%
\bibitem [{\citenamefont {Boyd}(2020)}]{BOYD20201}%
  \BibitemOpen
  \bibfield  {author} {\bibinfo {author} {\bibfnamefont {R.~W.}\ \bibnamefont
  {Boyd}},\ }\bibfield  {title} {\bibinfo {title} {Chapter 1 - the nonlinear
  optical susceptibility},\ }in\ \href
  {https://doi.org/https://doi.org/10.1016/B978-0-12-811002-7.00010-2} {\emph
  {\bibinfo {booktitle} {Nonlinear Optics (Fourth Edition)}}},\ \bibinfo
  {editor} {edited by\ \bibinfo {editor} {\bibfnamefont {R.~W.}\ \bibnamefont
  {Boyd}}}\ (\bibinfo  {publisher} {Academic Press},\ \bibinfo {year} {2020})\
  \bibinfo {edition} {fourth edition}\ ed.,\ pp.\ \bibinfo {pages}
  {1--64}\BibitemShut {NoStop}%
\bibitem [{\citenamefont {McCann}\ and\ \citenamefont {Koshino}(2013)}]{mck}%
  \BibitemOpen
  \bibfield  {author} {\bibinfo {author} {\bibfnamefont {E.}~\bibnamefont
  {McCann}}\ and\ \bibinfo {author} {\bibfnamefont {M.}~\bibnamefont
  {Koshino}},\ }\bibfield  {title} {\bibinfo {title} {The electronic properties
  of bilayer graphene},\ }\href@noop {} {\bibfield  {journal} {\bibinfo
  {journal} {Rep. Prog. Phys.}\ }\textbf {\bibinfo {volume} {78}},\ \bibinfo
  {pages} {056503} (\bibinfo {year} {2013})}\BibitemShut {NoStop}%
\bibitem [{\citenamefont {M\"ucke}\ \emph {et~al.}(2002)\citenamefont
  {M\"ucke}, \citenamefont {Tritschler}, \citenamefont {Wegener}, \citenamefont
  {Morgner},\ and\ \citenamefont {K\"artner}}]{mt}%
  \BibitemOpen
  \bibfield  {author} {\bibinfo {author} {\bibfnamefont {O.~D.}\ \bibnamefont
  {M\"ucke}}, \bibinfo {author} {\bibfnamefont {T.}~\bibnamefont {Tritschler}},
  \bibinfo {author} {\bibfnamefont {M.}~\bibnamefont {Wegener}}, \bibinfo
  {author} {\bibfnamefont {U.}~\bibnamefont {Morgner}},\ and\ \bibinfo {author}
  {\bibfnamefont {F.~X.}\ \bibnamefont {K\"artner}},\ }\bibfield  {title}
  {\bibinfo {title} {Role of the carrier-envelope offset phase of few-cycle
  pulses in nonperturbative resonant nonlinear optics},\ }\href
  {https://doi.org/10.1103/PhysRevLett.89.127401} {\bibfield  {journal}
  {\bibinfo  {journal} {Phys. Rev. Lett.}\ }\textbf {\bibinfo {volume} {89}},\
  \bibinfo {pages} {127401} (\bibinfo {year} {2002})}\BibitemShut {NoStop}%
\bibitem [{\citenamefont {Tritschler}\ \emph {et~al.}(2003)\citenamefont
  {Tritschler}, \citenamefont {M\"ucke}, \citenamefont {Wegener}, \citenamefont
  {Morgner},\ and\ \citenamefont {K\"artner}}]{tmw}%
  \BibitemOpen
  \bibfield  {author} {\bibinfo {author} {\bibfnamefont {T.}~\bibnamefont
  {Tritschler}}, \bibinfo {author} {\bibfnamefont {O.~D.}\ \bibnamefont
  {M\"ucke}}, \bibinfo {author} {\bibfnamefont {M.}~\bibnamefont {Wegener}},
  \bibinfo {author} {\bibfnamefont {U.}~\bibnamefont {Morgner}},\ and\ \bibinfo
  {author} {\bibfnamefont {F.~X.}\ \bibnamefont {K\"artner}},\ }\bibfield
  {title} {\bibinfo {title} {Evidence for third-harmonic generation in disguise
  of second-harmonic generation in extreme nonlinear optics},\ }\href
  {https://doi.org/10.1103/PhysRevLett.90.217404} {\bibfield  {journal}
  {\bibinfo  {journal} {Phys. Rev. Lett.}\ }\textbf {\bibinfo {volume} {90}},\
  \bibinfo {pages} {217404} (\bibinfo {year} {2003})}\BibitemShut {NoStop}%
\end{thebibliography}
apsrev4-2.bst 2019-01-14 (MD) hand-edited version of apsrev4-1.bst

\end{document}